\documentclass[floatfix,aps,prb,preprint,superscriptaddress,showpacs]{revtex4-1}
\usepackage{amsmath,amssymb,graphicx}

\def\dds{dd\sigma} \def\ddp{dd\pi} \def\ddd{dd\delta}
\def\sds{sd\sigma} \def\sss{ss\sigma}
\def\pds{pd\sigma} \def\pdp{pd\pi}
\def\e{{\rm e}}

\def\DS{$d$-sc} \def\DF{$d$-fx}
\def\SDS{$sd$-sc} \def\SDF{$sd$-fx}
\def\azbcc{a_0^{\rm{bcc}}}

\def\rredoz{r_1} \def\rredcz{r_c}
\def\sss{ss\sigma} \def\sds{sd\sigma} 
\def\sps{sp\sigma}
\def\pps{pp\sigma} \def\ppp{pp\pi} \def\pds{pd\sigma} \def\pdp{pd\pi}
\def\dds{dd\sigma} \def\ddp{dd\pi} \def\ddd{dd\delta} 
\def\e{\varepsilon}
\def\ione{{\it i\/}} \def\ii{{\it ii\/}} \def\iii{{\it iii\/}}    
 \def\ea{{\it et al.}}  

\begin{document}

\title{Analysis of a carbon dimer bound to a vacancy in iron using density
  functional theory and a new tight binding model}

\author{A. T. Paxton}
\altaffiliation[Present address: ]{ National Physical
  Laboratory, Teddington, Middlesex, TW11 0LW, UK {\it and}
  Department of Physics, King's 
  College London, Strand, London WC2R 2LS, UK}
\email{Tony.Paxton@KCL.ac.uk}
\affiliation{Fraunhofer Institut f\"ur
  Werkstoffmechanik IWM, W\"ohlerstr.~11, 79108 Freiburg,
  Germany}
\affiliation{Atomistic Simulation Centre, School of Mathematics
  and Physics, Queen's University Belfast, Belfast BT7 1NN, UK}
\author{C. Els{\"a}sser}
\email{Christian.Elsaesser@iwm.fraunhofer.de}
\affiliation{Fraunhofer Institut f\"ur
  Werkstoffmechanik IWM, W\"ohlerstr.~11, 79108 Freiburg,
  Germany}
\affiliation{Karlsruher Institut f\"ur Technologie, Institut
  f\"ur Ausgewandte Materialen (IAM-ZBS),
  Kaiserstr.~12, 76131 Karlsruhe, Germany}

\date{\today}

\begin{abstract}
  Recent density functional theory (DFT) calculations by
  F\"orst~\ea\cite{Forst06} have predicted that vacancies in both low
  and high carbon steels have a carbon dimer bound to them. This is
  likely to change the thinking of metallurgists in the kinetics of
  the development of microstructures. While the notion of a C$_2$
  molecule bound to a vacancy in Fe will potentially assume a central
  importance in the atomistic modeling of steels, neither a recent
  tight binding (TB) model nor existing classical interatomic
  potentials can account for it. Here we present a new TB model for C
  in Fe, based on our earlier work for H~in Fe, which correctly
  predicts the structure and energetics of the C$_2$ dimer at a
  vacancy in Fe. Moreover the model is capable of dealing with both
  concentrated and dilute limits of carbon in both $\alpha$-Fe and
  $\gamma$-Fe as comparisons with DFT show. We use both DFT and TB to
  make a detailed analysis of the dimer and to come to an
  understanding as to what governs the choice of its curious
  orientation within the vacancy.
\end{abstract}


\maketitle

\section{Introduction}
\label{sec_Intro}

A great deal of progress over many years has been made in
understanding the {\it physics} of the bonding and electronic
structure in pure magnetic iron\cite{Pettifor95} and this has been
used with great effect to advance the generation of schemes from
density functional theory in the gradient corrected\cite{PBE} local
spin density approximation\cite{Poulsen76,Christensen88} (LSDA-GGA)
through the tight binding approximation\cite{Liu05,Paxton08} to
classical interatomic potentials to be used for atomistic simulations
by the {\it materials science} community.\cite{Dudarev05} Of course
the metallurgist is rarely interested in pure iron and so the
challenge to the physicist has been to extend the theory to include
interstitial carbon, which is the defining element whose presence
distinguishes steel from iron. It is only in the last decade that real
progress has been made; first with some very extensive LSDA-GGA
calculations,\cite{Domain01,Domain04,Forst06,Becquart07,Monasterio09,Kabir10}
second with the generation of (admittedly very complicated) classical
interatomic potentials based in the embedded atom
method,\cite{Becquart07,Lau07} and third, the subject of this paper,
by some recent semi empirical quantum mechanical schemes based in the
tight binding approximation. A particularly significant advance was
made recently by Hatcher, Madsen and Drautz\cite{Hatcher12} who
constructed a very simple orthogonal tight binding model for carbon
and iron using a minimal basis of C-$p$ and Fe-$d$ orbitals and a
local charge neutrality condition. This model is a natural basis for a
bond order potential,\cite{Drautz11} but we argue here that this basis
may be too small to capture some of the physics of carbon in
iron. Instead we introduce a new model based in our earlier work on
H~in Fe\cite{Paxton10} employing a larger, non orthogonal basis of
C-$p$, Fe-$d$, and C~and Fe-$s$ orbitals and treating charge transfer
self consistently {\it via} an adjustable ``Hubbard-$U$''
parameter.\cite{Finnis98} The structure of the paper is as follows. In
section~\ref{sec_description} we describe our new model for carbon in
iron and in section~\ref{sec_predictions} we demonstrate its
predictive power in both the concentrated (iron carbide) and dilute
impurity limits. In~\ref{sec_dimer} we focus on the carbon dimer bound
to a vacancy, taking in view the recent startling prediction from
LSDA-GGA that this is a predominant point defect in
steel.\cite{Forst06} Our discussion and conclusions are to be found in
sections~\ref{sec_discussion} and~\ref{sec_conclusions}.

\section{Description of the new model}
\label{sec_description}

We take the same approach as in our earlier work on H~in
Fe\cite{Paxton10} which is similar to that of
Hatcher~\ea\cite{Hatcher12} on C in Fe, namely to proceed from a given
model for pure Fe and generate a further parameterization for the
interstitial element. In contrast to Hatcher~\ea\cite{Hatcher12} we do
not use a direct projection of the LSDA-GGA Hamiltonian onto a tight
binding basis,\cite{Madsen11,Urban11} instead we employ a genetic
algorithm\cite{Schwefel93} to fit the parameters to a small set of
LSDA-GGA target data which enter an objective function, which is
minimized. As a consequence of employing the simplest tight binding
scheme, namely an orthogonal basis of only $d$-orbitals on the Fe
atoms and $p$-orbitals on the C~atoms, the model of
Hatcher~\ea\cite{Hatcher12} differs significantly from ours.  One
difference results from their underlying model for pure Fe which
includes an {\it attractive\/} bonding term in the total energy which
is environment dependent and which accounts for a significant fraction
of the total energy.\cite{Madsen11} This was intended as a surrogate
for the missing $s$-electrons, but the fact that this term is large
and {\it negative} is surprising as one expects the $s$-band to exert
a {\it positive} pressure.\cite{Pettifor77,Varenna,Paxton96} Another
difference is that in the minimal basis having only $p$-orbitals on
C~atoms the limit of pure carbon can only be approximately rendered
since it is the $sp$-hybridisation in carbon that leads to the rich
variety of single, double and triple bonds and the competition between
$sp^2$-bonded graphite and $sp^3$-bonded diamond. We will argue below
that carbon $sp$-hybridization plays a key role in the structural
stability of iron carbides and also in controlling the configuration
of the C$_2$ dimer at an Fe vacancy. Therefore in the current work we
employ a larger basis, namely $s$-~and $d$-orbitals on Fe and $s$-~and
$p$-orbitals on C~atoms from which we suppose that at the expense of
greater computational cost we have a physically better motivated
model. Moreover we use a {\it non orthogonal} basis, and as we argued
earlier\cite{Paxton10} we believe that this allows a more natural way
to include environment dependence in the bond energy.

The tight binding model that we present here is identical in
its mathematical form to those we developed
earlier.\cite{Paxton10} The functional forms of the bond and
overlap integrals are
$$
    h(r) = h_0\,e^{-qr} \hskip 24pt   s(r) = s_0\,e^{-qr}
$$
and we tabulate all parameter values $h_0$ and $s_0$ in
tables~\ref{tbl_intersite-parameters} (for Fe--Fe terms)
and~\ref{tbl_Fe-CH-intersite-parameters} (for Fe--C
interactions). The Fe--Fe and Fe--C pair potentials are
$$
     \phi(r) = B_1 \,e^{-p_1r} - B_2 \,e^{-p_2r}
$$
noting the sign, so that in tables~\ref{tbl_intersite-parameters}
and~\ref{tbl_Fe-CH-intersite-parameters} parameter values for
$B_1$ and $B_2$ are positive. For Fe--H the pair potential is
$$ 
\phi(r) = {{B_1}\over{r}} \,e^{-p_1r} 
$$

\begin{table}
  \caption{\label{tbl_intersite-parameters} Intersite bond integral
    and pair potential parameters for the Fe--Fe terms in our tight
    binding model. All quantities are in Rydberg atomic units
    (a.u.) except for the cut off radii, $r_1$ and $r_c$ which are
    in units of the equilibrium bcc lattice constant
    $\azbcc$=2.87\AA. We include models using both volume-scaled and
    fixed cut offs (see the text). These differ only in their pair
    potentials and are indicated as ``sc'' and ``fx'' respectively in
    the last four rows. Properties of pure Fe resulting from these four
    models are displayed in
    table~\ref{tbl_pure-Fe-properties}. Parameters for C~and H~in Fe in
    table~\ref{tbl_Fe-CH-intersite-parameters} and used in the remainder of this paper are associated
    with the fixed cut off $sd$-model (\SDF). We note that these
    parameters differ from those published earlier;\cite{Paxton10} first by
    correcting misprints in the decay constant $q$ in the $\sss$\ and
    $\sds$\ terms, second because we have moved the cut off $r_1$ from
    10\% larger to 10\% smaller than $\azbcc$. Third we now prefer
    fixed multiplicative to scaled augmentative cut offs. 
    See also the text. These differences are
    then reflected in slightly different calculated properties in
    table~\ref{tbl_pure-Fe-properties}. }
 
\bigskip

\newdimen\digitwidth
\setbox0=\hbox{\rm0}
\digitwidth=\wd0
\centerline{
\vbox{
\catcode`~=\active
\def~{\kern\digitwidth}
\def\tablerule{\noalign{\smallskip\hrule\smallskip}}
\def\doubletablerule{\noalign{\smallskip\hrule\vskip 1pt\hrule\smallskip}}
\hrule\vskip 1pt\hrule
\medskip
\halign{
  \hfil#\hfil &\quad  \hfil#\hfil &\quad
  \hfil#\hfil &\quad  \hfil#\hfil &\quad
  \hfil#\hfil &\quad  \hfil#\hfil &\quad
  \hfil#\hfil \cr
  $\sss$\span\omit & $\sds$\span\omit &  $\dds$ &
  $\ddp$ &  $\ddd$ \cr
  $h_0$  & $s_0$  & $h_0$  & $s_0$ & $h_0$ & $h_0$ & $h_0$ \cr
\doubletablerule
 --0.35 & 0.45 & --0.14067 & 0.5 & --2.4383 & 1.9972 & --0.90724 \cr
\multispan4 \hrulefill $q=0.3$ \hrulefill & \multispan3 \hrulefill $q=0.9$ \hrulefill \cr
\multispan4 \hrulefill $r_1=1.1$ \hrulefill & \multispan3 \hrulefill $r_1=0.9$ \hrulefill \cr
\multispan4 \hrulefill $r_c=2.0$ \hrulefill & \multispan3 \hrulefill $r_c=1.4$ \hrulefill \cr
\noalign{\bigskip}
\doubletablerule
 & $B_1$ & $p_1$ & $B_2$ & $p_2$ & $\rredoz$ & $\rredcz$ \cr
\tablerule
\SDF & 698.67  & 1.52~~~~ & 517.467 & 1.4576~~ & 0.9 &  1.4  \cr
\SDS & 665.60  & 1.40843~ & 536.800 & 1.36297~ & 0.9 &  1.4  \cr
\DF  & 683.1~  & 1.5376~~ & 459.5~~ & 1.4544~~ & 0.9 &  1.4  \cr
\DS  & 682.8~  & 1.5165~~ & 466.8~~ & 1.4350~~ & 0.9 &  1.4  \cr
      }
\medskip
\hrule\vskip 1pt\hrule
}
}
\end{table}

\begin{table}
\newdimen\digitwidth
\setbox0=\hbox{\rm0}
\digitwidth=\wd0
\vskip -12pt
\caption{\label{tbl_pure-Fe-properties} Properties of pure
  $\alpha$-Fe. Target elastic constants are taken from
  experimental low temperature data (see
  ref~[\onlinecite{Paxton10}]); theoretical hcp--bcc energy difference
  from our own LSDA-GGA calculations; 
  experimental vacancy formation
  and migration energies are from Seeger,\cite{Seeger98} while remaining
  LSDA-GGA data are from Domain and Bequart,\cite{Domain01} and
  Kabir~\ea\cite{Kabir10} We show four models in data columns
  1--4: canonical ($d$) with scaled (sc) and fixed (fx) cut offs
  and non orthogonal ($sd$) with scaled and fixed cut offs
  respectively.}  
\centerline{ \vbox{ \catcode`~=\active
    \def~{\kern\digitwidth}
    \def\tablerule{\noalign{\smallskip\hrule\smallskip}}
    \def\doubletablerule{\noalign{\smallskip\hrule\vskip
        1pt\hrule\smallskip}} \hrule\vskip 1pt\hrule
\medskip
\halign{
  #\hfil &\quad  \hfil#\hfil &\quad
  \hfil#\hfil &\quad  \hfil#\hfil &\quad
  \hfil#\hfil &\quad  \hfil#\hfil &\quad  \hfil#\hfil \cr
 & \DS & \DF & \SDS & \SDF & target & \cr
\tablerule
$K$ (Gpa)        & 161 & 174 &  185 & 192 &  168  & (expt.)  \cr
$C'$ (GPa)       & ~50 & ~50 &  ~55 & 45  &  53   & (expt.) \cr
$c_{44}$ (Gpa)   & 118 & 117 &  106 & 100 & 122   & (expt.)  \cr
$\Delta E_{{\rm {hcp-bcc}}}$ (mRy) & ~~8 & ~~6 & ~~6 & ~~6 &  15
 & (LSDA-GGA) \cr
\noalign{\smallskip}
$H_{\rm Vac.}^{\rm F}$ (eV) & 2.0~ && 1.6~  & & 1.61--1.75 &  (expt.) \cr
                            &      &&       & &  2.0  & (LSDA-GGA) \cr
\noalign{\smallskip}
$H_{\rm Vac.}^{\rm M}$ (eV) & 1.16 && 0.81 & & 1.12--1.34   & (expt.)\cr
                            &      &&       & &  0.65--0.75 &  (LSDA-GGA) \cr
}
\medskip
\hrule\vskip 1pt\hrule
}
}
\end{table}

We take a rather sophisticated approach to cutting off the spacial
dependence of these interactions. We require proper energy
conservation in molecular dynamics and cannot allow discontinuities in
second derivatives of bond integral or pair potential
functions. Previously\cite{Paxton10} we implemented the cut off by
{\it augmenting} (that is, replacing) the function with a polynomial
of degree five within $r_1<r<r_c$ whose coefficients are chosen so as
to match the function continuously and differentiably to its {\it
  value} at $r_1$ and to {\it zero} at $r_c$. We chose $r_1=1.1\azbcc$
and $r_c=1.4\azbcc$, where $\azbcc = 2.87\hbox{~\AA}$ is the lattice
constant of $\alpha$-Fe, so that functions are cut off to zero between
second and third neighbors of the bcc lattice. Subsequent
improvements were introduced after making two
observations.\cite{Pashov12} (\ione) A smoother effect can be achieved
using a {\it multiplicative} cut off; that is, to multiply the
function by a polynomial of degree five whose value is one at $r_1$
and zero at $r_c$ and whose coefficients again ensure that the
function is everywhere continuous up to the second derivative.  (\ii)
Because the multiplicative cut off ``inherits'' the shape of the
function near $r_1$ better than the augmentative cut off, we found
that we could move $r_1$ back to $0.9\azbcc$ and achieve a smoother
function overall. A second difference compared to our earlier
work\cite{Paxton10} is that there we employed a volume dependent
cutoff, whereas now we prefer to use a cut off that is fixed. Because
of this small modification it is necessary to obtain slightly amended
pair potential parameters. We show these in
table~\ref{tbl_intersite-parameters}, and in
table~\ref{tbl_pure-Fe-properties} some predicted properties of pure
Fe using both the canonical $d$-band model for Fe and the non
orthogonal $sd$-model. Our canonical model can be read from
table~\ref{tbl_intersite-parameters} simply by ignoring those
parameters that don't enter the Hamiltonian. This model therefore
differs from the canonical model that we published
earlier.\cite{Paxton10} In fact it is worth pointing out that both
these canonical $d$-band models reproduce the vacancy formation and
migration energies better than our non orthogonal model; although its
$H_{\rm Vac.}^{\rm F}$ is outside the experimental range it is in
better agreement with published LSDA-GGA data. In our opinion the
simplest canonical model is very appropriate for pure Fe and we do not
see the need for the additional, attractive ``embedding potential''
introduced by Madsen~\ea\cite{Madsen11} The inclusion of the $s$-band
and non orthogonality in Fe is only necessary once hydrogen or first
row elements are included. The reason for this is that the valence
$s$-band from these elements lies typically below the Fe $3d$-bands;
orthogonality constraints in the concentrated limit then push the Fe
$4s$-band to above the Fermi level. In the dilute limit the electronic
structure has to differentiate between regions close to an impurity
and those far from it where the iron $4s$ local density of states
returns to its position in pure~Fe below the Fermi level. To account
properly for this effect the impurity and Fe $s$-bands cannot be
neglected. Of course in a minimal $pd$ basis for Fe--C tight binding
models or bond order potentials {\it both} $s$-bands are neglected
which is internally consistent, but these models cannot account for,
say, the carbon $sp$-hybridisation.

\begin{figure*}
  \caption{\label{fig_FeC-EV} (color online) Structural energy--volume
    curves for the four iron monocarbide phases, FeC, that were used
    in the fitting of the tight binding model. These show the heat of
    formation as a function of the atomic volume per Fe atom in both
    bcc ($\alpha$) and fcc ($\gamma$) Fe each containing C atoms in
    tetrahedral (TET) or octahedral (OCT) interstices. Note that the
    atomic volume, $\Omega_0$, of pure $\alpha$-Fe is 79.765~a.u. In
    this, and subsequent figures, GGA denotes the generalized gradient
    approximation to the LSDA.}
\begin{center}
\includegraphics[scale=0.8]{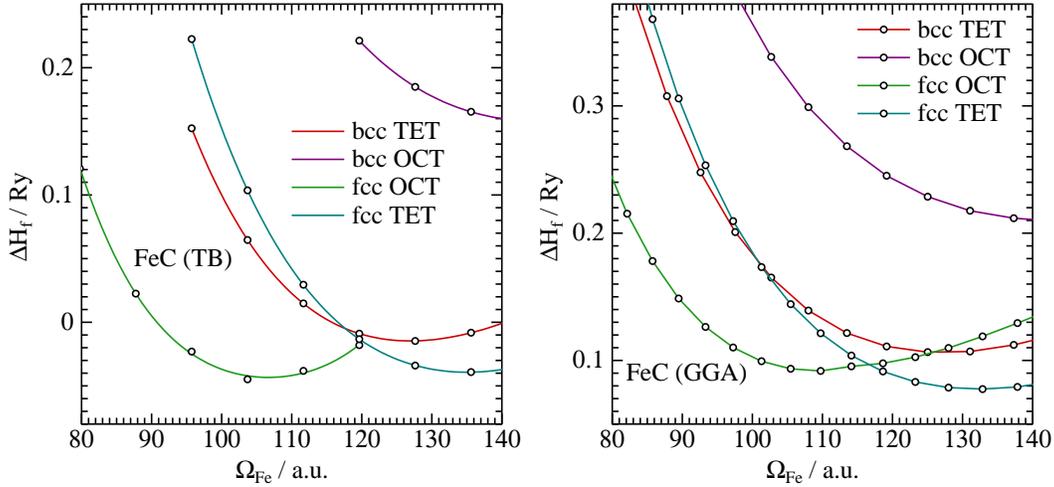}
\end{center}
\end{figure*}

\begin{table}
\caption{\label{tbl_on-site-parameters} On-site Hamiltonian
  matrix elements for tight binding models of Fe, H~and C. $U$ is
  the Hubbard-$U$ parameter and $J$ is the Stoner
  parameter.\cite{Christensen88,Paxton08} All quantities are in
  Rydberg atomic units (a.u.); $N_d$ is the number of
  $d$-electrons, which is an adjustable parameter in the
  canonical model.}

\bigskip

\newdimen\digitwidth
\setbox0=\hbox{\rm0}
\digitwidth=\wd0
\centerline{
\vbox{
\catcode`~=\active
\def~{\kern\digitwidth}
\def\tablerule{\noalign{\smallskip\hrule\smallskip}}
\def\doubletablerule{\noalign{\smallskip\hrule\vskip 1pt\hrule\smallskip}}
\hrule\vskip 1pt\hrule
\medskip
\halign{
  #\hfil &\quad  
  \hfil#\hfil &\quad  \hfil#\hfil &\quad
  \hfil#\hfil &\quad  \hfil#\hfil &\quad
  \hfil#\hfil \cr
 & $\e_s-\e_d^{\rm Fe}$ & $\e_p-\e_d^{\rm Fe}$ & $N_d$ & $U$ &  $J$ \cr
\tablerule
Fe-$d$~ &  &  &  7  &  & 0.05~ \cr
Fe-$sd$ & 0.15~ & & & 1.0~~ & 0.055 \cr
C       & --0.468~ & 0.083 & & 1.238 & 0 \cr
H       & --0.085~ &         & & 1.2~~ & 0 \cr
}
\medskip
\hrule\vskip 1pt\hrule
}
}
\end{table}

\begin{table}
\newdimen\digitwidth
\setbox0=\hbox{\rm0}
\digitwidth=\wd0
\vskip -12pt
\caption{\label{tbl_properties-monocarbides} Properties of iron
  monocarbides. These are the atomic volume of Fe, $\Omega_{\rm Fe}$, in
  FeC relative to $\Omega_0={(\azbcc)}^3/2$ in the four monocarbides obtained
  from bcc $\alpha$-Fe and fcc $\gamma$-Fe in which C~is in
  either tetrahedral or octahedral interstices. $\Delta E$ is the
  energy difference per formula unit in Ry compared to the fcc octahedral
  compound. Targets are taken from our LSDA-GGA calculations.
}

\bigskip

  \centerline{ \vbox{ \catcode`~=\active
    \def~{\kern\digitwidth}
    \def\tablerule{\noalign{\smallskip\hrule\smallskip}}
    \def\doubletablerule{\noalign{\smallskip\hrule\vskip
        1pt\hrule\smallskip}} \hrule\vskip 1pt\hrule
\medskip
\halign{
  \hfil#\hfil &\quad  \hfil#\hfil &\quad
  \hfil#\hfil &\quad  \hfil#\hfil &\quad
  \hfil#\hfil &\quad  \hfil#\hfil &\quad
  \hfil#\hfil &\quad  \hfil#\hfil \cr
 & \multispan2 $\alpha$-tet  & \multispan2 $\alpha$-oct  & \multispan2 $\gamma$-tet  & $\gamma$-oct \cr
 & $\Omega_{\rm Fe}/\Omega_0$ & $\Delta E$ & $\Omega_{\rm Fe}/\Omega_0$ & $\Delta E$ & $\Omega_{\rm Fe}/\Omega_0$ & $\Delta E$ & $\Omega_{\rm Fe}/\Omega_0$ \cr
\tablerule
TB      & 1.591 & 0.030 & 1.832 & 0.202 & 1.690 & ~0.005  & 1.327\cr
target  & 1.549 & 0.020 & 1.788 & 0.147 & 1.613 & --0.010 & 1.339\cr
}
\medskip
\hrule\vskip 1pt\hrule
}
}
\end{table}

\subsection{The on-site carbon and Fe--C parameters}
\label{subsec_FeC}

Our approach to finding carbon on-site energy parameters,
Fe--C Hamiltonian matrix elements and pair potential parameters
is to fit these to just seven target data (a refinement was
done later to improve the model in the dilute limit, see
section~\ref{sec_dilute}). These data are taken from LSDA-GGA
calculations and illustrated in figure~\ref{fig_FeC-EV} which
shows the heat of formation of four compounds having the
stoichiometry FeC. These are either bcc $\alpha$-Fe or fcc
$\gamma$-Fe with C interstitials in tetrahedral or octahedral
sites. In the $\gamma$-Fe case these are identical to the
zincblende and rocksalt crystal structures. We fitted our
parameters to the four equilibrium volumes and three energy
differences. The outcomes of the fitting are shown to the left of
figure~\ref{fig_FeC-EV} and in
table~\ref{tbl_properties-monocarbides}. The resulting parameter
values are tabulated in tables~\ref{tbl_intersite-parameters},
~\ref{tbl_on-site-parameters},
and~\ref{tbl_Fe-CH-intersite-parameters}. For comparison and for
completeness we also show parameters of our earlier model for
hydrogen in Fe.\cite{Paxton10}

It is important to make some comments about the energy--volume curves
in figure~\ref{fig_FeC-EV}, also in relation to the equivalent data
for the hydrogen interstitial.\cite{Paxton10} First, in the bcc
structure both C~and H~prefer the tetrahedral site in the concentrated
limit of FeC and FeH, and this remains the preferred site for H~into
the dilute limit. In contrast carbon occupies the octahedral sites in
both ferritic and austenitic steel. In $\alpha$-Fe, this is achieved
at the expense of a local tetragonal distortion of the lattice so as
to drive apart the two apical Fe atoms in the irregular octahedron of
the underlying bcc lattice. This is only possible if the C is
sufficiently dilute, certainly more dilute than the stoichiometry
Fe$_4$C, as we will see below, and in fact the crossover is around
Fe$_{16}$C.\cite{Jiang03} Second, in the fcc structure it is certainly
striking that according to LSDA-GGA FeC adopts the zincblende
structure rather than the rocksalt structure, albeit at an expanded
volume, as the tetrahedral interstice is much smaller than the
octahedral. This is contrary to the behavior of hydrogen, even though
its atomic radius is evidently smaller. Again there is a crossover
towards the dilute limit where C prefers the octahedral site in
$\gamma$-Fe.\cite{Jiang03} We expect that the competition between the
two sites in FeC is driven by the $sp$-hybridization which will be
maximal in the four fold coordinated tetrahedral site, whereas the six
fold octahedral site offers a bonding environment favorable to the
90$^{\circ}$ bond angles of unhybridised $p$-orbitals. Therefore it is
surprising that the $pd$-basis model\cite{Hatcher12} reproduces
this result correctly. We expect that this arises from the freedom of
employing long ranged C--C interactions in that model.\cite{Hatcher12}
Conversely we take the canonical point of view that C--C interactions
should not extend beyond the first neighbor distance in
diamond,\cite{Xu92} as it is known that longer range terms do not
improve tight binding models for diamond structure $sp$-bonded
elements.\cite{Harrison80,Paxton87}

\begin{table}
\caption{\label{tbl_Fe-CH-intersite-parameters} Intersite bond
  integrals and pair potential parameters 
  for the Fe--C and Fe--H terms.  All quantities are in
  Rydberg atomic units (a.u.) except for the cut off radii, $r_1$
  and $r_c$ which are in units of the equilibrium bcc lattice
  constant $\azbcc=2.87$\AA. This corrects two misprints regarding the
  $\sss$ and $\sds$ bond integrals for Fe--H in table~V of
  ref~[\onlinecite{Paxton10}]. }

\bigskip

\centerline{
\vbox{
\catcode`~=\active
\def~{\kern\digitwidth}
\def\tablerule{\noalign{\smallskip\hrule\smallskip}}
\def\doubletablerule{\noalign{\smallskip\hrule\vskip 1pt\hrule\smallskip}}
\hrule\vskip 1pt\hrule
\medskip
\halign{
  \hfil#\hfil &\hskip 3pt  \hfil#\hfil &\hskip 3pt
  \hfil#\hfil &\hskip 3pt  \hfil#\hfil &\hskip 3pt
  \hfil#\hfil &\hskip 3pt  \hfil#\hfil &\hskip 3pt
  \hfil#\hfil &\hskip 3pt  \hfil#\hfil &\hskip 3pt
  \hfil#\hfil &\hskip 3pt  \hfil#\hfil &\hskip 3pt
  \hfil#\hfil \cr
& $\sss$\span\omit & $\sps$\span\omit &
  $\sds$\span\omit & $\pds$\span\omit & $\pdp$\span\omit \cr
&  $h_0$ & $s_0$ & $h_0$ & $s_0$ & $h_0$ & $s_0$ & $h_0$ & $s_0$ & $h_0$ & $s_0$ \cr
&    $q$  &  $q$  &  $q$  &  $q$  &  $q$  &  $q$  &  $q$  &  $q$  &  $q$  &  $q$  \cr
& $\rredoz$ & $\rredoz$ & $\rredoz$ & $\rredoz$ & $\rredoz$ & $\rredoz$ & $\rredoz$ & $\rredoz$ & $\rredoz$ & $\rredoz$ \cr
& $\rredcz$ & $\rredcz$ & $\rredcz$ & $\rredcz$ & $\rredcz$ & $\rredcz$ & $\rredcz$ & $\rredcz$ & $\rredcz$ & $\rredcz$ \cr
\tablerule
 Fe--C & --1.7712 & 0.38434 & 3.9546 & --0.59202 & --0.17549 & 0.10283 & --1.2300 & 0.32895 & 0.88500 & --0.37025\cr
 &      ~0.56548 & 0.30106  & 0.76024 & 0.39114 & ~0.30249 & 0.34080 & ~0.64362 & 0.30636 & 0.66529 & ~0.45518 \cr
 &       \span \hrulefill 0.528 \hrulefill  &  \span \hrulefill 0.611 \hrulefill & 
         \multispan6 \hrulefill 0.595 \hrulefill \cr
 &       \span \hrulefill 1.790 \hrulefill  & \span \hrulefill 1.644 \hrulefill & 
         \multispan6 \hrulefill 1.674\hrulefill  \cr
Fe--H & --1.0935~~ & 0.26587 & & & --0.40748 & 0.21988 & & & & \cr
      & 0.77628    & 0.28633 & & &  ~0.45450 & 0.47301 & & & & \cr
      & 0.8        & 0.8     & & &   0.8     &  0.8    & & & & \cr
      & 2          & 2       & & &   2       &  2      & & & & \cr
\noalign{\medskip}
\tablerule
&& & $B_1$ & $p_1$ &  $B_2$ & $p_2$ & $\rredoz$ & $\rredcz$ & &\cr
\tablerule
&&Fe--C & 771.190 & 2.3962 & 19.325 & 1.5555 & 0.50071 & 1.5070 && \cr
&&Fe--H & 299.563 & 2.69225 &&& 0.75 & 0.95 && \cr
      }
\medskip
\hrule\vskip 1pt\hrule
}
}
\end{table}

\subsection{The C--C parameterization}
\label{subsec_CC}

\begin{figure*}
  \caption{\label{fig_Fe2C-EV} (color online) Energy volume curves for
    compounds with stoichiometry Fe$_2$C, comparing the predictions of
    our model with results of our LSDA-GGA calculations. These
    compounds have either bcc ($\alpha$) or fcc ($\gamma$) iron
    lattices with C placed at interstitial positions in tetrahedral
    (t), octahedral (o), in the bcc case the saddle point (s) along
    the $\langle 110\rangle$ direction and, in the fcc case, the
    saddle points (s) along the $\langle 111\rangle$ direction and (d)
    along the $\langle 110\rangle$ direction. This latter site is
    midway between two nearest neighbor Fe atoms and so is expected
    to have a high energy; on the other hand as is known from LSDA-GGA
    calculations\cite{Jiang03} and as our model also predicts, this
    site is along the diffusion path in $\gamma$-Fe.  The alternative
    path for diffusion (adopted by~H) {\it via} an intermediate
    tetrahedral site has higher energy. This is discussed in
    section~\ref{sec_dilute}. }
\begin{center}
\includegraphics[scale=0.8]{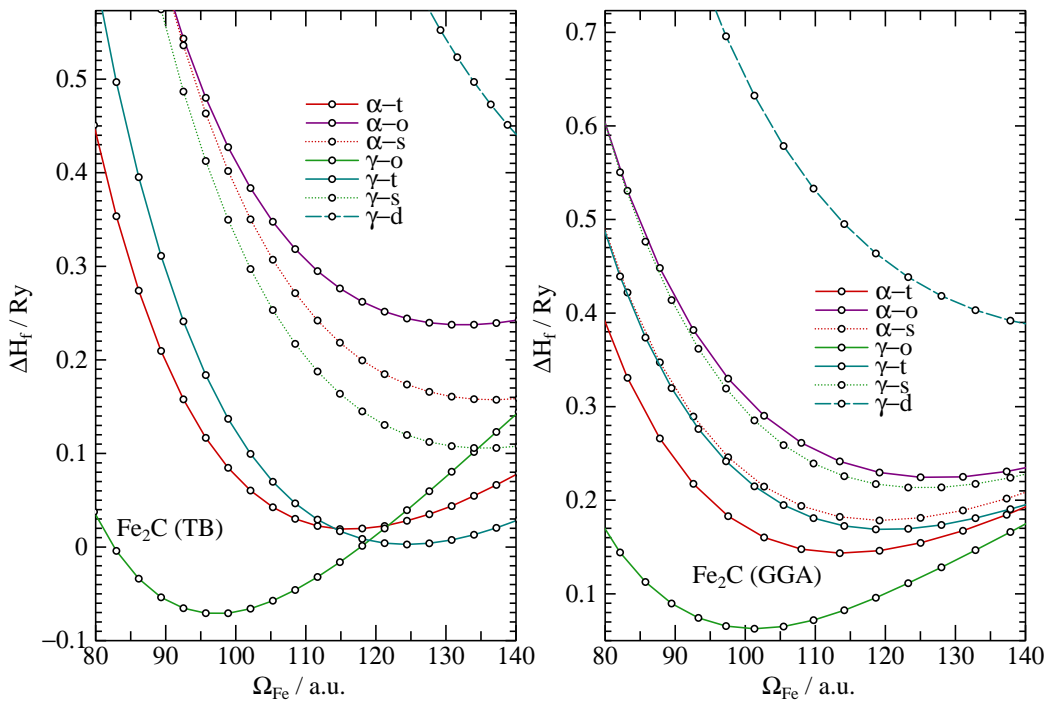}
\end{center}
\end{figure*}

\begin{figure*}
  \caption{\label{fig_Fe3C-EV} (color online) Energy volume curves for
    compounds with stoichiometry Fe$_3$C, comparing the predictions of
    our model with results of our LSDA-GGA calculations. Here we show
    both interstitial, and substitutional putative phases all of which
    have large positive heats of formation and are hence predicted not
    to exist. The first four in the column of labels represent bcc and
    fcc supercells containing one vacancy and a C atom at either a
    neighboring tetrahedral (t) or octahedral (o) site. The next four
    are substitutional phases labeled using their {\it
      Strukturbericht} designations. Of greatest interest are the
    remaining three Fe$_3$C compounds: ``WC'' labels a hypothetical
    high energy structure similar to a simple hexagonal tungsten
    carbide like phase (see the text); the only phases predicted to
    exist thermodynamically are the $\epsilon$ carbide and the
    $\theta$ carbide, or cementite phase. It is interesting to note
    that the LSDA-GGA predicts these both to have a {\it positive}
    heat of formation, which requires further investigation, since
    they are both known to exist ubiquitously in the microstructures
    of steels.}
\begin{center}
\includegraphics[scale=0.8]{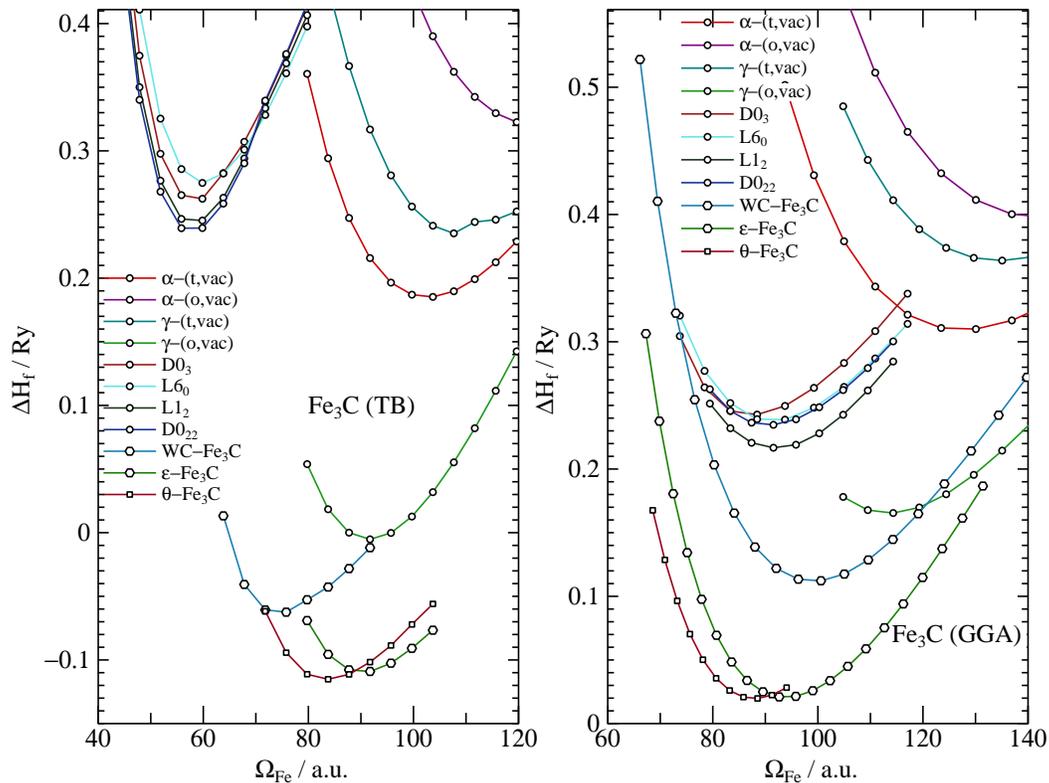}
\end{center}
\end{figure*}

\begin{figure*}
\caption{\label{fig_Fe4C-EV} (color online) Energy volume curves
  for compounds with stoichiometry Fe$_4$C, comparing the
  predictions of our tight binding model with our LSDA-GGA
  calculations. Here we show a variety of supercells in bcc
  ($\alpha$) and fcc ($\gamma$) crystal structures, each in face
  centered cubic (fcc), body centered tetragonal (bct) and simple
  cubic (sc) supercell settings. In each of these cases, a C atom
  is placed at one of the interstitial sites, using the same
  labeling as in figure~\ref{fig_Fe2C-EV}. As expected from the
  results of figure~\ref{fig_FeC-EV} and seen also in the data in
  figure~\ref{fig_Fe2C-EV} only the $\gamma$-Fe with octahedral
  carbon have reasonably low heats of formation.}
\begin{center}
\includegraphics[scale=0.8]{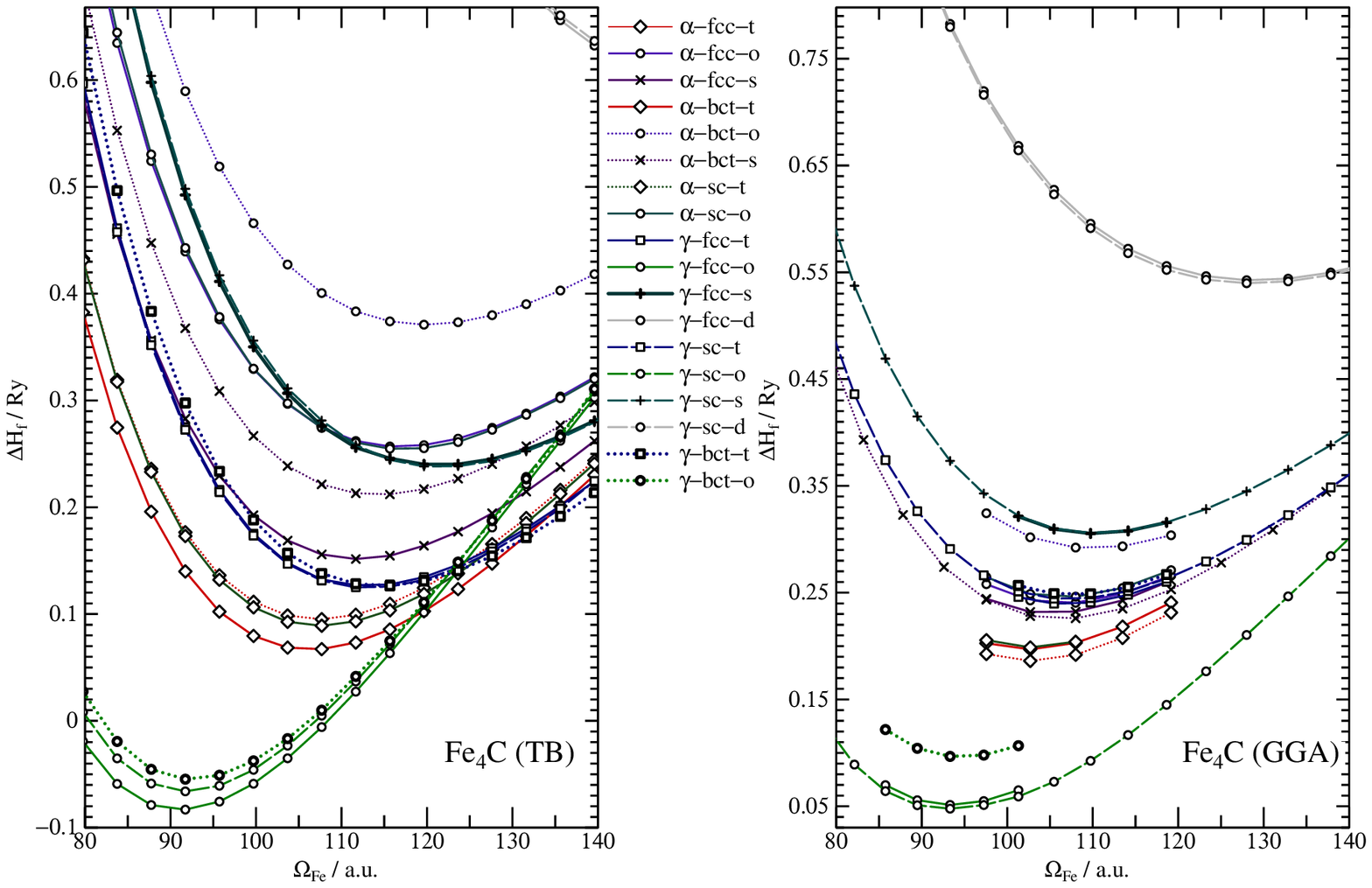}
\end{center}
\end{figure*}

Taking the view that carbon--carbon interactions are to be curtailed
beyond the usual definition of the chemical bond lengths of
1.2--1.5~\AA, none of the tests that we will apply in
section~\ref{sec_predictions} will require us to specify the C--C bond
integrals or pair potential. There is however one notable exception
which we will be discussing in greater detail below. This is the
observation from LSDA-GGA calculations\cite{Domain04} that two carbon
atoms bound to a monovacancy in Fe will form a ``dimer molecule''
whose bond length is 1.44~\AA. In our model we have the freedom to
choose our C--C interactions at will since they do not affect any of
the results in the concentrated limit. It would be desirable if an
existing model for diamond could be adopted without modification and
we have used a tight binding Hamiltonian for diamond from
Harrison\cite{Harrison80} with parameters adapted by Xu~\ea\cite{Xu92}
(Our model is essentially that of Harrison in terms of the scaling of
the bond integrals and pair potential with bond length.  We take over
the bond integrals at the equilibrium volume in diamond from Xu~\ea\
but we do not adopt their scaling.)  In this way for the C--C
parameters we use a simple power law model, namely
$$
    h(r) = h_0\,r^{-2} \hskip 6pt \hbox{,} \hskip 18pt \phi(r) = B_1\,r^{-4}
$$
with (in Rydberg a.u.)
$$
    h_0^{\sss}=-3.734\qquad h_0^{\sps}=3.510\qquad
    h_0^{\pps}=4.107\qquad h_0^{\ppp}=-1.157\qquad
$$
$B_1=50$~a.u.~leads to the correct lattice constant {\it and} bulk
modulus in diamond carbon. Unfortunately that choice of $B_1$ does not
quite reproduce the energy and C--C bond length of the C$_2$ molecule
bound to the Fe monovacancy. Therefore we employ $B_1=43$~a.u.~which
leads to about an 8\% error in the diamond lattice constant. We have
used the same value of $B_1$ to calculate the total energy of diamond
which is the quantity we have used in
figures~\ref{fig_FeC-EV}--\ref{fig_Fe4C-EV} to determine the heat of
formation of Fe--C compounds from elemental $\alpha$-Fe and
diamond~C. This has the consequence that the tight binding theory
consistently overestimates $-\Delta H_f$ by 0.16~Ry (see for example
figure~\ref{fig_FeC-EV}); if we use the value $B_1=50$~a.u.~the
agreement with the LSDA-GGA is rather better. To keep the C--C
interactions to within the first neighbors as expected in diamond and
in hydrocarbons, we apply a fixed multiplicative cut-off at $r_1=0.6\azbcc$
and $r_c=\azbcc$.

\section{Predictions of the new model}
\label{sec_predictions}

In this section we examine to what extent our model reproduces
some previously published or our calculated LSDA-GGA data.

\subsection{The concentrated limit}
\label{sec_concentrated}

We first compare tight binding with LSDA-GGA for a range of mostly
fictitious Fe--C compounds with stoichiometries Fe$_2$C, Fe$_3$C, and
Fe$_4$C in figures~\ref{fig_Fe2C-EV}--\ref{fig_Fe4C-EV}. The LSDA-GGA
calculations were done by means of the mixed-basis pseudopotential
(MBPP) method.\cite{Els90,Lecherman02} The PBE-GGA
exchange and correlation functional,\cite{PBE} optimally smooth
norm conserving pseudopotentials\cite{Vanderbilt85} for Fe and C, {\bf
  k}-points which are equivalent to $8\times8\times8$ Chadi-Cohen
meshes for cubic structures, and a Gaussian broadening by 0.2~eV were
employed. The mixed basis consisted of plane waves up to a maximum
kinetic energy of 340~eV and atom-centered basis functions with
$d$-symmetry for Fe atoms and with $p$-symmetry for C atoms which are
confined to spheres with radii of 1.19\AA\ and 0.66\AA\
respectively. The broad agreement between tight binding and LSDA-GGA
is excellent and indeed in almost all instances the ordering in energy
of the phases is correctly reproduced.

\begin{table}
\newdimen\digitwidth
\setbox0=\hbox{\rm0}
\digitwidth=\wd0
\caption{\label{tbl_epsilon-theta-carbides} Calculated lattice
  parameters of hexagonal $\epsilon$~Fe$_3$C and orthorhombic
  $\theta$~Fe$_3$C (cementite), compared to LSDA-GGA calculations
  and experimental measurements reported by Jang~\ea\cite{Jang09,Jang10}
  Lattice parameters $a$, $b$ and $c$ are in \AA. The last column
  shows the predicted equilibrium volume compared to experiment.}
\centerline{ \vbox{ \catcode`~=\active \def~{\kern\digitwidth}
    \def\tablerule{\noalign{\smallskip\hrule\smallskip}}
    \def\doubletablerule{\noalign{\smallskip\hrule\vskip
        1pt\hrule\smallskip}} \hrule\vskip 1pt\hrule
\medskip
\halign{
  \hfil#\hfil &\quad  \hfil#\hfil &\quad \hfil#\hfil &\quad
  \hfil#\hfil &\quad  \hfil#\hfil &\quad
  \hfil#\hfil &\quad  \hfil#\hfil &\quad\hfil#\hfil \cr
 & & $a$ & $b$ & $c$ & $b/a$ & $c/a$ & $V/V_{\rm exp}$ \cr
\tablerule
$\theta$-Fe$_3$C & TB    & 4.95 & 6.79 & 4.42 & 1.37 & 0.89 & 0.96 \cr
                 & LSDA-GGA   & 5.13 & 6.65 & 4.46 & 1.30 & 0.86 & 0.98 \cr
                 & exp. & 5.09 & 6.74 & 4.52 & 1.32 & 0.89 & \cr
$\epsilon$-Fe$_3$C & TB & 4.63 &  & 8.64 & & 1.87  & 0.93 \cr
                 & LSDA-GGA  & 4.74 &  & 8.63 & & 1.82  & 0.98 \cr
                 & exp. & 4.77 &  & 8.71 & & 1.83  & \cr
}
\medskip
\hrule\vskip 1pt\hrule
}
}
\end{table}

We will focus most closely on the stoichiometry Fe$_3$C,
figure~\ref{fig_Fe3C-EV}, which is the most significant composition in
materials science due to the ubiquitous occurence of cementite in the
microstructures of steels. It is also notable that a significant
weakness in the tight binding model emerges here when trying to
describe the hypothetical {\it substitutional} phases D0$_3$, L6$_0$,
L1$_2$, and D0$_{22}$. This is not so surprising since this bonding
environment is very different from the {\it interstitial} phases and
fortunately the substitutional phases are not of particularly great
interest. It is on the other hand very gratifying that the tight
binding model reproduces with great fidelity the $\epsilon$~and
$\theta$~iron carbide phases. At the same time the ordering of the
unfavorable simple hexagonal tungsten carbide like structure is very
well rendered; this hypothetical structure is obtained from the
hexagonal close packed $\epsilon$-Fe$_3$C by rotating alternate Fe
layers about the $c$-axis by 60$^{\circ}$ and increasing the axial
ratio by $\sqrt{3/2}$. To emphasise the suitability of the tight
binding approximation in the modeling of steel microstructures, we
show in table~\ref{tbl_epsilon-theta-carbides} a detailed comparison
of the calculated crystal lattice parameters of the important phases,
$\epsilon$~and $\theta$~iron carbide, with experimental data.

\subsection{The dilute limit}
\label{sec_dilute}

Of equal or even greater interest is the behavior of carbon in the
dilute limit. In the case of hydrogen in Fe we could claim a success
in that a model fitted in the most concentrated stoichiometry
transfers very well into the dilute limit.\cite{Paxton10} Carbon in Fe
has been more problematic and subsequent to the fitting described in
section~\ref{subsec_FeC} it was necessary to make further genetic
optimizations of the Fe--C parameters in order to render correctly the
migration energy and the binding of C to a monovacancy---the
quantities $H^{\rm M_{\alpha}}_{\rm C}$ and $E_B$(1) in the first two
data columns of table~\ref{tbl_dilute-properties}.

\begin{table}
\newdimen\digitwidth
\setbox0=\hbox{\rm0}
\digitwidth=\wd0
\vskip -12pt
\caption{\label{tbl_dilute-properties} Properties of C in
  Fe in the dilute limit. $H^{\rm M_{\alpha}}_{\rm C}$ is the migration
  energy of the C atom, equal to the energy difference between C
  in tetrahedral and octahedral interstices in bcc $\alpha$-Fe. $E_B$(1) and $E_B$(2k) are the
  binding energies of one and two C atoms to a vacancy
  (see ref~[\onlinecite{Becquart07}], table~5).
  The final two columns are data for C~in fcc
  $\gamma$-Fe and show the migration energies $H^{\rm
    M_{\gamma}}_{\rm C}$(tet) and $H^{\rm M_{\gamma}}_{\rm C}$(d)
  of C between two octahedral sites {\it via} a tetrahedral site and
  the ``d-saddle'' site respectively. All energies are in eV. The
  first line shows LSDA-GGA data taken from
  refs~[\onlinecite{Domain01}], [\onlinecite{Jiang03}] and
  [\onlinecite{Lau07}]. The second line shows results from our TB model.
}

\bigskip

\centerline{ \vbox{ \catcode`~=\active \def~{\kern\digitwidth}
    \def\tablerule{\noalign{\smallskip\hrule\smallskip}}
    \def\doubletablerule{\noalign{\smallskip\hrule\vskip
        1pt\hrule\smallskip}} \hrule\vskip 1pt\hrule
\medskip
\halign{
  \hfil#\hfil &\quad  \hfil#\hfil &\quad
  \hfil#\hfil &\quad  \hfil#\hfil &\quad
  \hfil#\hfil &\quad  \hfil#\hfil \cr
 & $H^{\rm M_{\alpha}}_{\rm C}$ & $E_B$(1) &
 $E_B$(2k) &  $H^{\rm M_{\gamma}}_{\rm C}$(tet) &
 $H^{\rm M_{\gamma}}_{\rm C}$(d) \cr
\tablerule
LSDA-GGA    & 0.87   & 0.47  & 1.50   & 1.48 & 1.00\cr
TB          & 0.81   & 0.35  & 1.55   & 2.11 & 0.63  \cr
}
\medskip
\hrule\vskip 1pt\hrule
}
}
\end{table}

\begin{table}
\newdimen\digitwidth
\setbox0=\hbox{\rm0}
\digitwidth=\wd0
\vskip -12pt
\caption{\label{tbl_dimer-properties} Comparative energies and bond
  lengths of four possible configurations of two carbon atoms bound to
  one vacancy in bcc $\alpha$-Fe. We use the designations of
  refs~[\onlinecite{Domain01}] and [\onlinecite{Lau07}]. The
  structures relaxed using TB are illustrated in
  figure~\ref{fig_Dimer}. The first and second rows show
  results from LSDA-GGA calculations and the third and fourth those from the TB
  model. The latter correctly predicts the ordering in energy (these
  are shown relative to the ``k'' ground state energy) and the bond
  length. The ``$\langle 110\rangle$'' configuration has the dimer
  centered at the vacant Fe
  site and orientated along a $\langle 110\rangle$ direction. Energy differences
  $\Delta E$ are in
  eV and bond lengths $d$ in~\AA. The numbers in parentheses are
  energy differences calculated using the Fe atom positions of the
  relaxed equivalent structure, with both C~atoms removed; these numbers
  allow a comparison of the host lattice distortions
  accompanying the formation of the dimer--vacancy complex. 
}

\bigskip

\centerline{ \vbox{ \catcode`~=\active \def~{\kern\digitwidth}
    \def\tablerule{\noalign{\smallskip\hrule\smallskip}}
    \def\doubletablerule{\noalign{\smallskip\hrule\vskip
        1pt\hrule\smallskip}} \hrule\vskip 1pt\hrule
\medskip
\halign{
  \hfil#\hfil &\qquad  \hfil#\hfil &\quad
  \hfil#\hfil &\qquad  \hfil#\hfil &\quad
  \hfil#\hfil &\qquad  \hfil#\hfil &\quad
  \hfil#\hfil &\qquad  \hfil#\hfil &\quad
  \hfil#\hfil \cr
 & ``j''\span\omit & ``$\langle 100\rangle$''\span\omit & ``k''\span\omit &
 ``$\langle 110\rangle$''\span\omit \cr
\tablerule
 & $\Delta E$ & $d$  & $\Delta E$ & $d$  & $\Delta E$ & $d$  & $\Delta E$ & $d$ \cr
LSDA-GGA    & 0.37 & 2.57 & 0.11 & 1.46 & 0 & 1.43 & 0.06 & 1.43 \cr
            & (0.07) &   & (--0.05) &   &(0)&    & (0.04) &      \cr
TB          & 1.10 & 2.70 & 0.13 & 1.46 & 0 & 1.44 & 1.23 & 1.40 \cr
            & (0.49)  &  & (0.16)  &    &(0)&    & (0.66) &      \cr
}
\medskip
\hrule\vskip 1pt\hrule
}
}
\end{table}

Our main results are presented in tables~\ref{tbl_dilute-properties}
and~\ref{tbl_dimer-properties}. We have constructed cubic $4\times 4
\times 4$ supercells for $\alpha$-Fe and $\gamma$-Fe in order to
study, in particular, the energetics of the monovacancy in Fe and its
binding to carbon interstitials. As in the case of hydrogen, the
impurity does not occupy a vacant Fe lattice site, as is clear from
figure~\ref{fig_Fe3C-EV} which shows a large positive heat of
formation for the four substitutional phases considered. Instead
(again, as does hydrogen) carbon occupies a position close to its
preferred interstitial site, in this case the octahedral interstice,
at one of the cube faces bounding the vacancy. We follow the
definitions employed by Becquart~\ea\cite{Becquart07} such that the
binding energy of one or more interstitials to a vacancy is the
difference in energy between that number of interstitials and the
vacancy occupying separate, non interacting sites, and the
interstitials bound to the vacancy.  In this way, we have, from
calculations based on a 128-atom supercell of pure Fe,
\begin{equation*}
E_B(1)=-\big(E(\hbox{Fe}_{127}\hbox{C})+E(\hbox{Fe}_{128})\big)
       +\big(E(\hbox{Fe}_{127})+E(\hbox{Fe}_{128}\hbox{C})\big)
\end{equation*}
and
\begin{equation*}
E_B(2)=-\big(E(\hbox{Fe}_{127}\hbox{C}_{2})+2E(\hbox{Fe}_{128})\big)
        +\big(E(\hbox{Fe}_{127})+2E(\hbox{Fe}_{128}\hbox{C})\big)
\label{eq_EB2}
\end{equation*}
where the signs are employed such that a positive binding energy
implies a preference for the two C~atoms to bind at a vacancy compared
to the vacancy and two C~interstitials being widely separated. The
total energies $E$ involved are calculated by relaxing supercells
containing the numbers of atoms indicated in parentheses.

We therefore show in table~\ref{tbl_dilute-properties} $E_B(1)$, the
binding energy of a single C atom to a monovacancy, and $E_B$(2k)
(using the designations of Becquart~\ea\cite{Becquart07}) the binding
energy of two C~atoms to a vacancy.

We have also calculated migration energies of carbon in $\alpha$-Fe and
$\gamma$-Fe using static relaxations and also the nudged elastic band
method.\cite{Henkelman00} TB describes these correctly in both phases
of Fe as seen in the data columns 1, 4, and 5 in
table~\ref{tbl_dilute-properties}. In particular our TB model confirms
the LSDA-GGA result that the diffusion path of C~in $\gamma$-Fe is
{\it not} as one might suppose mediated by a ``double'' hop {\it via}
a neighboring tetrahedral site as for H~in $\gamma$-Fe, but the
carbon atom actually takes a direct route forcing itself through the
bond center of two nearest neighbor Fe atoms at the $\langle
110\rangle$~(d) saddle point. This is surprising in view of the high
energy of the $\gamma$-d crystal structures in
figures~\ref{fig_Fe3C-EV} and~\ref{fig_Fe4C-EV}. However this tight
binding prediction agrees with LSDA-GGA results.\cite{Jiang04}

\subsection{The carbon dimer at the vacancy}
\label{sec_dimer}

Several authors have made LSDA-GGA calculations for a carbon dimer
bound to a vacancy in $\alpha$-Fe and described a number of possible
atomic structures.\cite{Domain04,Becquart07,Lau07} We consider four
here. If the dimer is orientated along a $\langle 100\rangle$
direction, then if the carbon atoms remain close to their original
octahedral sites at opposite faces of the cube bounding the vacancy,
this configuration is designated ``j'' by
Becquart~\ea\cite{Becquart07} or ``OO'' by Lau~\ea\cite{Lau07} This is
a {\it local minimum} in the energy and in our TB model the two carbon
atoms are outside their range of interaction in the Hamiltonian (see
section~\ref{subsec_CC}). The energy is lowered if the two carbon
atoms approach each other along the $\langle 100\rangle$ direction and
form a dimer bond, having a bond length of 1.46\AA\ which is very
close to that in diamond and the C--C single bond in molecules. This
configuration is denoted ``$\langle 100\rangle$'' by
Lau~\ea\cite{Lau07} but was not considered in earlier
work.\cite{Domain04,Becquart07} This is not yet the global minimum
energy for the dimer which is achieved by orientating the dimer along
a $\langle 011\rangle$ direction with the two C~atoms close to
octahedral positions, a situation denoted ``k'' by
Becquart~\ea\cite{Becquart07} or ``AO'' by Lau~\ea\cite{Lau07}
The latter authors describe a further configuration, ``$\langle
011\rangle$'' in which the dimer bond is centered at the vacant site
and orientated parallel to ``k''. This is of slightly higher energy
than ``k''. The four configurations are are illustrated in
figure~\ref{fig_Dimer}.  In table~\ref{tbl_dimer-properties} we
compare predictions of the TB model with our own LSDA-GGA
results.\cite{VASP_Note}\nocite{Kresse96,Kresse99} The only serious
discrepancy is that the TB model overestimates the energies of
the ``j'' and ``$\langle 011\rangle$'' configurations with respect to 
LSDA-GGA.

\section{Discussion}
\label{sec_discussion}

\begin{figure*}
  \caption{\label{fig_Dimer} (color online) Atomic structures of four
    configurations for a carbon dimer bound to a monovacency in
    $\alpha$-Fe. Structures shown are (a)
    ``j''[\onlinecite{Becquart07}], or ``OO''[\onlinecite{Lau07}]; (b)
    ``$\langle 100\rangle$''[\onlinecite{Lau07}]; (c)``$\langle
    110\rangle$''[\onlinecite{Lau07}]; (d)
    ``k''[\onlinecite{Becquart07}] or ``AO''[\onlinecite{Lau07}], the
    global minimum for this configuration.\cite{Lau07} The relaxed
    structures displayed here are obtained with our TB model; Fe--C
    bond lengths shown are 3.73\AA\ in ``$\langle 100\rangle$'' and
    3.65\AA\ in ``k''}
\begin{center}
\includegraphics[scale=1.4]{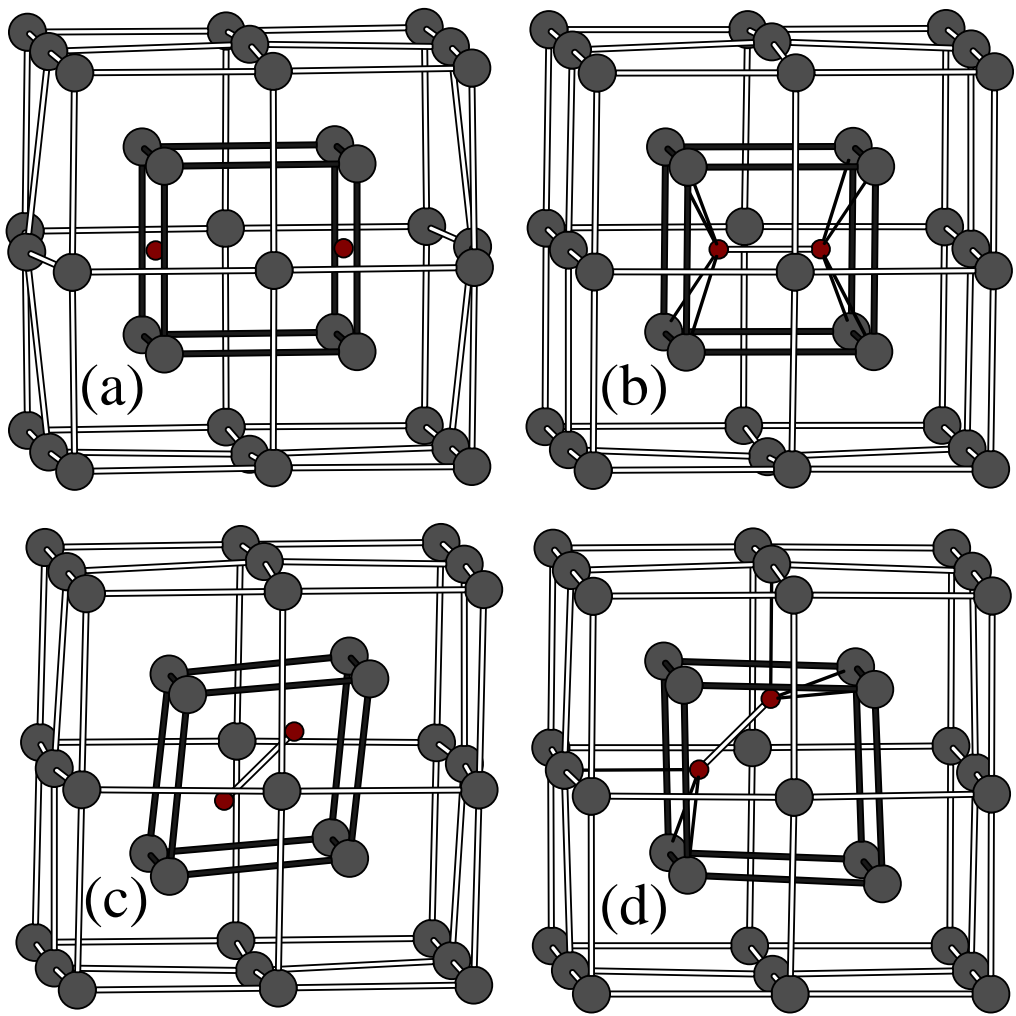}
\end{center}
\end{figure*}

F\"orst~\ea\cite{Forst06} made a very thorough study using LSDA-GGA of
point defect complex energetics and found the remarkable result that
under the conditions normally encountered in a steel, effectively {\it
  all Fe vacancies have a C$_2$ dimer bound to them} as illustrated in
figure~\ref{fig_Dimer}. Moreover, by just a few hundredths of an eV,
the dimer prefers to be orientated along a $\langle 110\rangle$
direction. It would be of great interest to determine whether this
phenomenon can be confirmed experimentally, possibly by internal
friction measurements. It is notable that a similar prediction was
made using DFT concerning dimerization of boron in copper, a
prediction that is consistent with thermodynamic assessment.\cite{Lozovoi08}

One can interpret the competition between the four configurations
illustrated in figure~\ref{fig_Dimer} in terms of certain notional
contributions to the total energy. These are (\ione)~the formation of
a C--C covalent bond, (\ii)~the coordination of the carbon atoms to
neighboring Fe atoms, and (\iii)~the accompanying distortion of the
host Fe lattice containing a vacancy. (\ione)~It is surprising that a
C--C bond having the same length as in pure carbon is formed in view
of the metallic electron gas destroying the single bond order; the
bond length inside the metal is the same as in the pure carbon or
hydrocarbon, but the bond energy is about ten times smaller. (\ii)~The
Fe--C coordination goes a long way to explain the stability of the
most stable configuration ``k'' in which the carbon dimer makes three
bonds of equal length (3.65\AA) to Fe atoms, thus forming an
``ethane'' molecule in which the Fe atoms take the part of H~atoms
(see figure~\ref{fig_Dimer}(d)). With the dimer orientated along
$\langle 100\rangle$ each carbon atom makes four bonds (3.73\AA~long)
to neighboring Fe atoms (figure~\ref{fig_Dimer}(b)). We take it that
this is less favorable owing to carbon preferring a four fold
coordination. Configurations ``j'' and ``$\langle 110\rangle$'' both
display a planar configuration of Fe--C bonds; in ``j'' the carbon is
bonded to four Fe atoms in the plane of a cube face of the bcc
lattice, in ``$\langle 110\rangle$'' each carbon is bonded to two Fe
atoms. (\iii)~In table~\ref{tbl_dimer-properties} we give in
parentheses values of the calculated total energies of the four
configurations, {\it having removed the two carbon atoms} and leaving
the Fe atoms in their positions. This is intended to examine the
elastic distortion energy accompanying the introduction of the dimer
into the vacancy. In the LSDA-GGA these distortion energies are rather
small and in fact ``$\langle 100\rangle$'' has a slightly lower value
than ``k''. However the preferred four fold coordination of the carbon
atoms in ``k'' is able to compensate for the increased distortion
energy. Whereas the TB model correctly predicts ``k'' to be the global
minimum, the quantitative comparison with LSDA-GGA is rather poor. We
suppose that this is a consequence of the small basis set of TB and
more limited self consistency. Thus TB {\it overestimates} effects
based on covalent bonding and lattice distortions because the
Hamiltonian does not have the degrees of freedom of LSDA-GGA to find a
lower energy in the case of an unfavorable structure. This is well
illustrated in the case of ``$\langle 110\rangle$'' in
figure~\ref{fig_Dimer}(c). Clearly there is a large distortion of the
bcc cubic unit cell surrounding the vacancy. The LSDA-GGA given the
constraint of this distortion in the {\it atomic} structure is yet
able to find a low energy {\it electronic} structure that can
accommodate the constraint. The TB finds this much more difficult.

\section{Conclusions}
\label{sec_conclusions}

The TB model presented here is intended as a physically better
motivated and more transferable scheme as compared to the recently
published orthogonal $pd$-model\cite{Hatcher12} or to existing
classical interatomic potentials.  Its transferability has been
demonstrated using tests in both concentrated and dilute limits, for
example successfully predicting the structure and energetics of
cementite (section~\ref{sec_concentrated}) and the migration path of
C~in $\gamma$-Fe (section~\ref{sec_dilute}).  It may be thought that
this migration path particularly would expose the need for environment
dependence in an empirical model\cite{Tang96,Haas98} 
Our model, using only two-center parameters, is able to deal with this
through the use of an overlap matrix between non orthogonal Fe and C
orbitals.\cite{Pettifor00,Urban11}

Our model also correctly describes the structure and energetics of the
carbon dimer bound to a vacancy in $\alpha$-Fe---a defect that is
expected to take a central importance following the predictions of
F\"orst~\ea\cite{Forst06} Apart from a large overestimation of the
energy of the ``$\langle 011\rangle$'' dimer, our TB model properly
orders the structures and predicts ``k'' to have the lowest energy
although we were forced to modify an existing simple model for carbon
in order to achieve the correct C$_2$ bond length (see
section~\ref{subsec_CC}). It is notable that published classical
potentials\cite{Becquart07} and the minimal basis TB
model\cite{Hatcher12} cannot reproduce the stability of the carbon
dimer. An exception is the recent classical potential of
Lau~\ea\cite{Lau07} although this model greatly {\it overestimates}
the binding energy of the $\langle 011\rangle$ dimer.

In view of the apparent significance of carbon dimers existing in the
microstructure of steel and their possible interactions with
hydrogen\cite{Monasterio09} it is now a matter of importance that
plausible and efficient quantum mechanical models are produced. From
this point of view the present work assumes a particular significance.

\section*{Acknowledgments}

It is a pleasure to thank Charlotte Bequart for a helpful
correspondence.  This work was undertaken as a part of the project
MultiHy of the European Union's 7th Framework Programme ({\tt
  MultiHy.eu}).  Financial support from the German Federal Ministry
for Education and Research (BMBF) to the Fraunhofer IWM for
C.~E. (Grant number 02NUK009C) is gratefully acknowledged.

\def\JPCM{J.~Phys.:~Condens.~Matter.}
\def\PRB{Phys.~Rev.~B}
\def\PRL{Phys.~Rev.~Lett.}

%


\begin{thebibliography}{44}%
\makeatletter
\providecommand \@ifxundefined [1]{%
 \@ifx{#1\undefined}
}%
\providecommand \@ifnum [1]{%
 \ifnum #1\expandafter \@firstoftwo
 \else \expandafter \@secondoftwo
 \fi
}%
\providecommand \@ifx [1]{%
 \ifx #1\expandafter \@firstoftwo
 \else \expandafter \@secondoftwo
 \fi
}%
\providecommand \natexlab [1]{#1}%
\providecommand \enquote  [1]{``#1''}%
\providecommand \bibnamefont  [1]{#1}%
\providecommand \bibfnamefont [1]{#1}%
\providecommand \citenamefont [1]{#1}%
\providecommand \href@noop [0]{\@secondoftwo}%
\providecommand \href [0]{\begingroup \@sanitize@url \@href}%
\providecommand \@href[1]{\@@startlink{#1}\@@href}%
\providecommand \@@href[1]{\endgroup#1\@@endlink}%
\providecommand \@sanitize@url [0]{\catcode `\\12\catcode `\$12\catcode
  `\&12\catcode `\#12\catcode `\^12\catcode `\_12\catcode `\%12\relax}%
\providecommand \@@startlink[1]{}%
\providecommand \@@endlink[0]{}%
\providecommand \url  [0]{\begingroup\@sanitize@url \@url }%
\providecommand \@url [1]{\endgroup\@href {#1}{\urlprefix }}%
\providecommand \urlprefix  [0]{URL }%
\providecommand \Eprint [0]{\href }%
\providecommand \doibase [0]{http://dx.doi.org/}%
\providecommand \selectlanguage [0]{\@gobble}%
\providecommand \bibinfo  [0]{\@secondoftwo}%
\providecommand \bibfield  [0]{\@secondoftwo}%
\providecommand \translation [1]{[#1]}%
\providecommand \BibitemOpen [0]{}%
\providecommand \bibitemStop [0]{}%
\providecommand \bibitemNoStop [0]{.\EOS\space}%
\providecommand \EOS [0]{\spacefactor3000\relax}%
\providecommand \BibitemShut  [1]{\csname bibitem#1\endcsname}%
\let\auto@bib@innerbib\@empty
\bibitem [{\citenamefont {F\"orst}\ \emph {et~al.}(2006)\citenamefont
  {F\"orst}, \citenamefont {Slycke}, \citenamefont {Van~Vliet},\ and\
  \citenamefont {Yip}}]{Forst06}%
  \BibitemOpen
  \bibfield  {author} {\bibinfo {author} {\bibfnamefont {C.~J.}\ \bibnamefont
  {F\"orst}}, \bibinfo {author} {\bibfnamefont {J.}~\bibnamefont {Slycke}},
  \bibinfo {author} {\bibfnamefont {K.~J.}\ \bibnamefont {Van~Vliet}}, \ and\
  \bibinfo {author} {\bibfnamefont {S.}~\bibnamefont {Yip}},\ }\href@noop {}
  {\bibfield  {journal} {\bibinfo  {journal} {Phys. Rev. Lett.}\ }\textbf
  {\bibinfo {volume} {96}},\ \bibinfo {pages} {175501} (\bibinfo {year}
  {2006})}\BibitemShut {NoStop}%
\bibitem [{\citenamefont {Pettifor}(1995)}]{Pettifor95}%
  \BibitemOpen
  \bibfield  {author} {\bibinfo {author} {\bibfnamefont {D.~G.}\ \bibnamefont
  {Pettifor}},\ }\href@noop {} {\emph {\bibinfo {title} {Bonding and structure
  of molecules and solids}}}\ (\bibinfo  {publisher} {Oxford University
  Press},\ \bibinfo {address} {Oxford},\ \bibinfo {year} {1995})\BibitemShut
  {NoStop}%
\bibitem [{\citenamefont {Perdew}\ \emph {et~al.}(1997)\citenamefont {Perdew},
  \citenamefont {Burke},\ and\ \citenamefont {Ernzerhof}}]{PBE}%
  \BibitemOpen
  \bibfield  {author} {\bibinfo {author} {\bibfnamefont {J.~P.}\ \bibnamefont
  {Perdew}}, \bibinfo {author} {\bibfnamefont {K.}~\bibnamefont {Burke}}, \
  and\ \bibinfo {author} {\bibfnamefont {M.}~\bibnamefont {Ernzerhof}},\
  }\href@noop {} {\bibfield  {journal} {\bibinfo  {journal} {Phys. Rev. Lett.}\
  }\textbf {\bibinfo {volume} {78}},\ \bibinfo {pages} {1396} (\bibinfo {year}
  {1997})}\BibitemShut {NoStop}%
\bibitem [{\citenamefont {Poulsen}\ \emph {et~al.}(1976)\citenamefont
  {Poulsen}, \citenamefont {Koll{\'a}r},\ and\ \citenamefont
  {Andersen}}]{Poulsen76}%
  \BibitemOpen
  \bibfield  {author} {\bibinfo {author} {\bibfnamefont {U.~K.}\ \bibnamefont
  {Poulsen}}, \bibinfo {author} {\bibfnamefont {J.}~\bibnamefont {Koll{\'a}r}},
  \ and\ \bibinfo {author} {\bibfnamefont {O.~K.}\ \bibnamefont {Andersen}},\
  }\href@noop {} {\bibfield  {journal} {\bibinfo  {journal} {J.~Phys.~F: Metal
  Phys.}\ }\textbf {\bibinfo {volume} {9}},\ \bibinfo {pages} {L241} (\bibinfo
  {year} {1976})}\BibitemShut {NoStop}%
\bibitem [{\citenamefont {Christensen}\ \emph {et~al.}(1988)\citenamefont
  {Christensen}, \citenamefont {Gunnarsson}, \citenamefont {Jepsen},\ and\
  \citenamefont {Andersen}}]{Christensen88}%
  \BibitemOpen
  \bibfield  {author} {\bibinfo {author} {\bibfnamefont {N.~E.}\ \bibnamefont
  {Christensen}}, \bibinfo {author} {\bibfnamefont {O.}~\bibnamefont
  {Gunnarsson}}, \bibinfo {author} {\bibfnamefont {O.}~\bibnamefont {Jepsen}},
  \ and\ \bibinfo {author} {\bibfnamefont {O.~K.}\ \bibnamefont {Andersen}},\
  }\href@noop {} {\bibfield  {journal} {\bibinfo  {journal} {J.~de
  Phys.~Colloque~C8}\ }\textbf {\bibinfo {volume} {49}},\ \bibinfo {pages} {17}
  (\bibinfo {year} {1988})}\BibitemShut {NoStop}%
\bibitem [{\citenamefont {Liu}\ \emph {et~al.}(2005)\citenamefont {Liu},
  \citenamefont {Nguyen-Manh}, \citenamefont {Liu},\ and\ \citenamefont
  {Pettifor}}]{Liu05}%
  \BibitemOpen
  \bibfield  {author} {\bibinfo {author} {\bibfnamefont {G.}~\bibnamefont
  {Liu}}, \bibinfo {author} {\bibfnamefont {D.}~\bibnamefont {Nguyen-Manh}},
  \bibinfo {author} {\bibfnamefont {B.-G.}\ \bibnamefont {Liu}}, \ and\
  \bibinfo {author} {\bibfnamefont {D.~G.}\ \bibnamefont {Pettifor}},\
  }\href@noop {} {\bibfield  {journal} {\bibinfo  {journal} {\PRB}\ }\textbf
  {\bibinfo {volume} {71}},\ \bibinfo {pages} {174115} (\bibinfo {year}
  {2005})}\BibitemShut {NoStop}%
\bibitem [{\citenamefont {Paxton}\ and\ \citenamefont
  {Finnis}(2008)}]{Paxton08}%
  \BibitemOpen
  \bibfield  {author} {\bibinfo {author} {\bibfnamefont {A.~T.}\ \bibnamefont
  {Paxton}}\ and\ \bibinfo {author} {\bibfnamefont {M.~W.}\ \bibnamefont
  {Finnis}},\ }\href@noop {} {\bibfield  {journal} {\bibinfo  {journal} {Phys.
  Rev. B}\ }\textbf {\bibinfo {volume} {77}},\ \bibinfo {eid} {024428}
  (\bibinfo {year} {2008})}\BibitemShut {NoStop}%
\bibitem [{\citenamefont {Dudarev}\ and\ \citenamefont
  {Derlet}(2005)}]{Dudarev05}%
  \BibitemOpen
  \bibfield  {author} {\bibinfo {author} {\bibfnamefont {S.~L.}\ \bibnamefont
  {Dudarev}}\ and\ \bibinfo {author} {\bibfnamefont {P.~M.}\ \bibnamefont
  {Derlet}},\ }\href@noop {} {\bibfield  {journal} {\bibinfo  {journal}
  {\JPCM}\ }\textbf {\bibinfo {volume} {17}},\ \bibinfo {pages} {7097}
  (\bibinfo {year} {2005})}\BibitemShut {NoStop}%
\bibitem [{\citenamefont {Domain}\ and\ \citenamefont
  {Becquart}(2001)}]{Domain01}%
  \BibitemOpen
  \bibfield  {author} {\bibinfo {author} {\bibfnamefont {C.}~\bibnamefont
  {Domain}}\ and\ \bibinfo {author} {\bibfnamefont {C.~S.}\ \bibnamefont
  {Becquart}},\ }\href@noop {} {\bibfield  {journal} {\bibinfo  {journal}
  {Phys. Rev. B}\ }\textbf {\bibinfo {volume} {65}},\ \bibinfo {pages} {024103}
  (\bibinfo {year} {2001})}\BibitemShut {NoStop}%
\bibitem [{\citenamefont {Domain}\ \emph {et~al.}(2004)\citenamefont {Domain},
  \citenamefont {Becquart},\ and\ \citenamefont {Foct}}]{Domain04}%
  \BibitemOpen
  \bibfield  {author} {\bibinfo {author} {\bibfnamefont {C.}~\bibnamefont
  {Domain}}, \bibinfo {author} {\bibfnamefont {C.~S.}\ \bibnamefont
  {Becquart}}, \ and\ \bibinfo {author} {\bibfnamefont {J.}~\bibnamefont
  {Foct}},\ }\href@noop {} {\bibfield  {journal} {\bibinfo  {journal} {Phys.
  Rev. B}\ }\textbf {\bibinfo {volume} {69}},\ \bibinfo {pages} {144112}
  (\bibinfo {year} {2004})}\BibitemShut {NoStop}%
\bibitem [{\citenamefont {Becquart}\ \emph {et~al.}(2007)\citenamefont
  {Becquart}, \citenamefont {Raulot}, \citenamefont {Bencteux}, \citenamefont
  {Domain}, \citenamefont {Perez}, \citenamefont {Garruchet},\ and\
  \citenamefont {Nguyen}}]{Becquart07}%
  \BibitemOpen
  \bibfield  {author} {\bibinfo {author} {\bibfnamefont {C.}~\bibnamefont
  {Becquart}}, \bibinfo {author} {\bibfnamefont {J.}~\bibnamefont {Raulot}},
  \bibinfo {author} {\bibfnamefont {G.}~\bibnamefont {Bencteux}}, \bibinfo
  {author} {\bibfnamefont {C.}~\bibnamefont {Domain}}, \bibinfo {author}
  {\bibfnamefont {M.}~\bibnamefont {Perez}}, \bibinfo {author} {\bibfnamefont
  {S.}~\bibnamefont {Garruchet}}, \ and\ \bibinfo {author} {\bibfnamefont
  {H.}~\bibnamefont {Nguyen}},\ }\href@noop {} {\bibfield  {journal} {\bibinfo
  {journal} {Computational Materials Science}\ }\textbf {\bibinfo {volume}
  {40}},\ \bibinfo {pages} {119 } (\bibinfo {year} {2007})}\BibitemShut
  {NoStop}%
\bibitem [{\citenamefont {Monasterio}\ \emph {et~al.}(2009)\citenamefont
  {Monasterio}, \citenamefont {Lau}, \citenamefont {Yip},\ and\ \citenamefont
  {Van~Vliet}}]{Monasterio09}%
  \BibitemOpen
  \bibfield  {author} {\bibinfo {author} {\bibfnamefont {P.~R.}\ \bibnamefont
  {Monasterio}}, \bibinfo {author} {\bibfnamefont {T.~T.}\ \bibnamefont {Lau}},
  \bibinfo {author} {\bibfnamefont {S.}~\bibnamefont {Yip}}, \ and\ \bibinfo
  {author} {\bibfnamefont {K.~J.}\ \bibnamefont {Van~Vliet}},\ }\href@noop {}
  {\bibfield  {journal} {\bibinfo  {journal} {Phys. Rev. Lett.}\ }\textbf
  {\bibinfo {volume} {103}},\ \bibinfo {pages} {085501} (\bibinfo {year}
  {2009})}\BibitemShut {NoStop}%
\bibitem [{\citenamefont {Kabir}\ \emph {et~al.}(2010)\citenamefont {Kabir},
  \citenamefont {Lau}, \citenamefont {Lin}, \citenamefont {Yip},\ and\
  \citenamefont {Van~Vliet}}]{Kabir10}%
  \BibitemOpen
  \bibfield  {author} {\bibinfo {author} {\bibfnamefont {M.}~\bibnamefont
  {Kabir}}, \bibinfo {author} {\bibfnamefont {T.~T.}\ \bibnamefont {Lau}},
  \bibinfo {author} {\bibfnamefont {X.}~\bibnamefont {Lin}}, \bibinfo {author}
  {\bibfnamefont {S.}~\bibnamefont {Yip}}, \ and\ \bibinfo {author}
  {\bibfnamefont {K.~J.}\ \bibnamefont {Van~Vliet}},\ }\href@noop {} {\bibfield
   {journal} {\bibinfo  {journal} {Phys. Rev. B}\ }\textbf {\bibinfo {volume}
  {82}},\ \bibinfo {pages} {134112} (\bibinfo {year} {2010})}\BibitemShut
  {NoStop}%
\bibitem [{\citenamefont {Lau}\ \emph {et~al.}(2007)\citenamefont {Lau},
  \citenamefont {F\"orst}, \citenamefont {Lin}, \citenamefont {Gale},
  \citenamefont {Yip},\ and\ \citenamefont {Van~Vliet}}]{Lau07}%
  \BibitemOpen
  \bibfield  {author} {\bibinfo {author} {\bibfnamefont {T.~T.}\ \bibnamefont
  {Lau}}, \bibinfo {author} {\bibfnamefont {C.~J.}\ \bibnamefont {F\"orst}},
  \bibinfo {author} {\bibfnamefont {X.}~\bibnamefont {Lin}}, \bibinfo {author}
  {\bibfnamefont {J.~D.}\ \bibnamefont {Gale}}, \bibinfo {author}
  {\bibfnamefont {S.}~\bibnamefont {Yip}}, \ and\ \bibinfo {author}
  {\bibfnamefont {K.~J.}\ \bibnamefont {Van~Vliet}},\ }\href@noop {} {\bibfield
   {journal} {\bibinfo  {journal} {Phys. Rev. Lett.}\ }\textbf {\bibinfo
  {volume} {98}},\ \bibinfo {pages} {215501} (\bibinfo {year}
  {2007})}\BibitemShut {NoStop}%
\bibitem [{\citenamefont {Hatcher}\ \emph {et~al.}(2012)\citenamefont
  {Hatcher}, \citenamefont {Madsen},\ and\ \citenamefont {Drautz}}]{Hatcher12}%
  \BibitemOpen
  \bibfield  {author} {\bibinfo {author} {\bibfnamefont {N.}~\bibnamefont
  {Hatcher}}, \bibinfo {author} {\bibfnamefont {G.~K.}\ \bibnamefont {Madsen}},
  \ and\ \bibinfo {author} {\bibfnamefont {R.}~\bibnamefont {Drautz}},\
  }\href@noop {} {\bibfield  {journal} {\bibinfo  {journal} {Phys. Rev. B}\
  }\textbf {\bibinfo {volume} {86}},\ \bibinfo {pages} {155115} (\bibinfo
  {year} {2012})}\BibitemShut {NoStop}%
\bibitem [{\citenamefont {Drautz}\ and\ \citenamefont
  {Pettifor}(2011)}]{Drautz11}%
  \BibitemOpen
  \bibfield  {author} {\bibinfo {author} {\bibfnamefont {R.}~\bibnamefont
  {Drautz}}\ and\ \bibinfo {author} {\bibfnamefont {D.~G.}\ \bibnamefont
  {Pettifor}},\ }\href@noop {} {\bibfield  {journal} {\bibinfo  {journal}
  {Phys. Rev. B}\ }\textbf {\bibinfo {volume} {84}},\ \bibinfo {pages} {214114}
  (\bibinfo {year} {2011})}\BibitemShut {NoStop}%
\bibitem [{\citenamefont {Paxton}\ and\ \citenamefont
  {Els\"asser}(2010)}]{Paxton10}%
  \BibitemOpen
  \bibfield  {author} {\bibinfo {author} {\bibfnamefont {A.~T.}\ \bibnamefont
  {Paxton}}\ and\ \bibinfo {author} {\bibfnamefont {C.}~\bibnamefont
  {Els\"asser}},\ }\href@noop {} {\bibfield  {journal} {\bibinfo  {journal}
  {Phys. Rev. B}\ }\textbf {\bibinfo {volume} {82}},\ \bibinfo {pages} {235125}
  (\bibinfo {year} {2010})}\BibitemShut {NoStop}%
\bibitem [{\citenamefont {Finnis}\ \emph {et~al.}(1998)\citenamefont {Finnis},
  \citenamefont {Paxton}, \citenamefont {Methfessel},\ and\ \citenamefont {van
  Schilfgaarde}}]{Finnis98}%
  \BibitemOpen
  \bibfield  {author} {\bibinfo {author} {\bibfnamefont {M.~W.}\ \bibnamefont
  {Finnis}}, \bibinfo {author} {\bibfnamefont {A.~T.}\ \bibnamefont {Paxton}},
  \bibinfo {author} {\bibfnamefont {M.}~\bibnamefont {Methfessel}}, \ and\
  \bibinfo {author} {\bibfnamefont {M.}~\bibnamefont {van Schilfgaarde}},\
  }\href@noop {} {\bibfield  {journal} {\bibinfo  {journal} {\PRL}\ }\textbf
  {\bibinfo {volume} {81}},\ \bibinfo {pages} {5149} (\bibinfo {year}
  {1998})}\BibitemShut {NoStop}%
\bibitem [{\citenamefont {Madsen}\ \emph {et~al.}(2011)\citenamefont {Madsen},
  \citenamefont {McEniry},\ and\ \citenamefont {Drautz}}]{Madsen11}%
  \BibitemOpen
  \bibfield  {author} {\bibinfo {author} {\bibfnamefont {G.~K.~H.}\
  \bibnamefont {Madsen}}, \bibinfo {author} {\bibfnamefont {E.~J.}\
  \bibnamefont {McEniry}}, \ and\ \bibinfo {author} {\bibfnamefont
  {R.}~\bibnamefont {Drautz}},\ }\href@noop {} {\bibfield  {journal} {\bibinfo
  {journal} {Phys. Rev. B}\ }\textbf {\bibinfo {volume} {83}},\ \bibinfo
  {pages} {184119} (\bibinfo {year} {2011})}\BibitemShut {NoStop}%
\bibitem [{\citenamefont {Urban}\ \emph {et~al.}(2011)\citenamefont {Urban},
  \citenamefont {Reese}, \citenamefont {Mrovec}, \citenamefont {Els\"asser},\
  and\ \citenamefont {Meyer}}]{Urban11}%
  \BibitemOpen
  \bibfield  {author} {\bibinfo {author} {\bibfnamefont {A.}~\bibnamefont
  {Urban}}, \bibinfo {author} {\bibfnamefont {M.}~\bibnamefont {Reese}},
  \bibinfo {author} {\bibfnamefont {M.}~\bibnamefont {Mrovec}}, \bibinfo
  {author} {\bibfnamefont {C.}~\bibnamefont {Els\"asser}}, \ and\ \bibinfo
  {author} {\bibfnamefont {B.}~\bibnamefont {Meyer}},\ }\href@noop {}
  {\bibfield  {journal} {\bibinfo  {journal} {Phys. Rev. B}\ }\textbf {\bibinfo
  {volume} {84}},\ \bibinfo {pages} {155119} (\bibinfo {year}
  {2011})}\BibitemShut {NoStop}%
\bibitem [{\citenamefont {Schwefel}(1993)}]{Schwefel93}%
  \BibitemOpen
  \bibfield  {author} {\bibinfo {author} {\bibfnamefont {H.-P.}\ \bibnamefont
  {Schwefel}},\ }\href@noop {} {\emph {\bibinfo {title} {Evolution and Optimum
  Seeking: The Sixth Generation}}}\ (\bibinfo  {publisher} {John Wiley},\
  \bibinfo {address} {New York},\ \bibinfo {year} {1993})\BibitemShut {NoStop}%
\bibitem [{\citenamefont {Pettifor}(1977)}]{Pettifor77}%
  \BibitemOpen
  \bibfield  {author} {\bibinfo {author} {\bibfnamefont {D.~G.}\ \bibnamefont
  {Pettifor}},\ }\href@noop {} {\bibfield  {journal} {\bibinfo  {journal}
  {Journal of Physics F: Metal Physics}\ }\textbf {\bibinfo {volume} {7}},\
  \bibinfo {pages} {613} (\bibinfo {year} {1977})}\BibitemShut {NoStop}%
\bibitem [{\citenamefont {Andersen}\ \emph {et~al.}(1985)\citenamefont
  {Andersen}, \citenamefont {Jepsen},\ and\ \citenamefont
  {Gl{\"o}tzel}}]{Varenna}%
  \BibitemOpen
  \bibfield  {author} {\bibinfo {author} {\bibfnamefont {O.~K.}\ \bibnamefont
  {Andersen}}, \bibinfo {author} {\bibfnamefont {O.}~\bibnamefont {Jepsen}}, \
  and\ \bibinfo {author} {\bibfnamefont {D.}~\bibnamefont {Gl{\"o}tzel}},\
  }\enquote {\bibinfo {title} {Highlights of condensed matter theory},}\ \
  (\bibinfo  {publisher} {North--Holland},\ \bibinfo {address} {New York},\
  \bibinfo {year} {1985})\ Chap.~\bibinfo {chapter} {3}\BibitemShut {NoStop}%
\bibitem [{\citenamefont {Paxton}(1996)}]{Paxton96}%
  \BibitemOpen
  \bibfield  {author} {\bibinfo {author} {\bibfnamefont {A.~T.}\ \bibnamefont
  {Paxton}},\ }\href@noop {} {\bibfield  {journal} {\bibinfo  {journal}
  {Journal of Physics D: Applied Physics}\ }\textbf {\bibinfo {volume} {29}},\
  \bibinfo {pages} {1689} (\bibinfo {year} {1996})}\BibitemShut {NoStop}%
\bibitem [{\citenamefont {Seeger}(1998)}]{Seeger98}%
  \BibitemOpen
  \bibfield  {author} {\bibinfo {author} {\bibfnamefont {A.}~\bibnamefont
  {Seeger}},\ }\href@noop {} {\bibfield  {journal} {\bibinfo  {journal} {phys.
  stat. sol. (a)}\ }\textbf {\bibinfo {volume} {167}},\ \bibinfo {pages} {289}
  (\bibinfo {year} {1998})}\BibitemShut {NoStop}%
\bibitem [{\citenamefont {Pashov}(2012)}]{Pashov12}%
  \BibitemOpen
  \bibfield  {author} {\bibinfo {author} {\bibfnamefont {D.}~\bibnamefont
  {Pashov}},\ }\emph {\bibinfo {title} {Electronic structure of certain
  Titanium-Aluminium superalloys: from first principles to Bond Order
  Potentials}},\ \href@noop {} {Ph.D. thesis},\ \bibinfo  {school} {Queen's
  University Belfast} (\bibinfo {year} {2012})\BibitemShut {NoStop}%
\bibitem [{\citenamefont {Jiang}\ and\ \citenamefont {Carter}(2003)}]{Jiang03}%
  \BibitemOpen
  \bibfield  {author} {\bibinfo {author} {\bibfnamefont {D.~E.}\ \bibnamefont
  {Jiang}}\ and\ \bibinfo {author} {\bibfnamefont {E.~A.}\ \bibnamefont
  {Carter}},\ }\href@noop {} {\bibfield  {journal} {\bibinfo  {journal} {Phys.
  Rev. B}\ }\textbf {\bibinfo {volume} {67}},\ \bibinfo {pages} {214103}
  (\bibinfo {year} {2003})}\BibitemShut {NoStop}%
\bibitem [{\citenamefont {Xu}\ \emph {et~al.}(1992)\citenamefont {Xu},
  \citenamefont {Wang}, \citenamefont {Chan},\ and\ \citenamefont {Ho}}]{Xu92}%
  \BibitemOpen
  \bibfield  {author} {\bibinfo {author} {\bibfnamefont {C.~H.}\ \bibnamefont
  {Xu}}, \bibinfo {author} {\bibfnamefont {C.~Z.}\ \bibnamefont {Wang}},
  \bibinfo {author} {\bibfnamefont {C.~T.}\ \bibnamefont {Chan}}, \ and\
  \bibinfo {author} {\bibfnamefont {K.~M.}\ \bibnamefont {Ho}},\ }\href@noop {}
  {\bibfield  {journal} {\bibinfo  {journal} {Journal of Physics: Condensed
  Matter}\ }\textbf {\bibinfo {volume} {4}},\ \bibinfo {pages} {6047} (\bibinfo
  {year} {1992})}\BibitemShut {NoStop}%
\bibitem [{\citenamefont {Harrison}(1980)}]{Harrison80}%
  \BibitemOpen
  \bibfield  {author} {\bibinfo {author} {\bibfnamefont {W.~A.}\ \bibnamefont
  {Harrison}},\ }\href@noop {} {\emph {\bibinfo {title} {Electronic structure
  and the properties of solids}}}\ (\bibinfo  {publisher} {W.~H.~Freeman},\
  \bibinfo {address} {San Francisco},\ \bibinfo {year} {1980})\BibitemShut
  {NoStop}%
\bibitem [{\citenamefont {Paxton}\ \emph {et~al.}(1987)\citenamefont {Paxton},
  \citenamefont {Sutton},\ and\ \citenamefont {Nex}}]{Paxton87}%
  \BibitemOpen
  \bibfield  {author} {\bibinfo {author} {\bibfnamefont {A.~T.}\ \bibnamefont
  {Paxton}}, \bibinfo {author} {\bibfnamefont {A.~P.}\ \bibnamefont {Sutton}},
  \ and\ \bibinfo {author} {\bibfnamefont {C.~M.~M.}\ \bibnamefont {Nex}},\
  }\href@noop {} {\bibfield  {journal} {\bibinfo  {journal} {Journal of Physics
  C: Solid State Physics}\ }\textbf {\bibinfo {volume} {20}},\ \bibinfo {pages}
  {L263} (\bibinfo {year} {1987})}\BibitemShut {NoStop}%
\bibitem [{\citenamefont {Els{\"a}sser}\ \emph {et~al.}(1990)\citenamefont
  {Els{\"a}sser}, \citenamefont {Takeuchi}, \citenamefont {Ho}, \citenamefont
  {Chan}, \citenamefont {Braun},\ and\ \citenamefont {F{\"a}hnle}}]{Els90}%
  \BibitemOpen
  \bibfield  {author} {\bibinfo {author} {\bibfnamefont {C.}~\bibnamefont
  {Els{\"a}sser}}, \bibinfo {author} {\bibfnamefont {N.}~\bibnamefont
  {Takeuchi}}, \bibinfo {author} {\bibfnamefont {K.~M.}\ \bibnamefont {Ho}},
  \bibinfo {author} {\bibfnamefont {C.~T.}\ \bibnamefont {Chan}}, \bibinfo
  {author} {\bibfnamefont {P.}~\bibnamefont {Braun}}, \ and\ \bibinfo {author}
  {\bibfnamefont {M.}~\bibnamefont {F{\"a}hnle}},\ }\href@noop {} {\bibfield
  {journal} {\bibinfo  {journal} {Journal of Physics: Condensed Matter}\
  }\textbf {\bibinfo {volume} {2}},\ \bibinfo {pages} {4371} (\bibinfo {year}
  {1990})}\BibitemShut {NoStop}%
\bibitem [{\citenamefont {Lechermann}\ \emph {et~al.}(2002)\citenamefont
  {Lechermann}, \citenamefont {Welsch}, \citenamefont {Els\"asser},
  \citenamefont {Ederer}, \citenamefont {F\"ahnle}, \citenamefont {Sanchez},\
  and\ \citenamefont {Meyer}}]{Lecherman02}%
  \BibitemOpen
  \bibfield  {author} {\bibinfo {author} {\bibfnamefont {F.}~\bibnamefont
  {Lechermann}}, \bibinfo {author} {\bibfnamefont {F.}~\bibnamefont {Welsch}},
  \bibinfo {author} {\bibfnamefont {C.}~\bibnamefont {Els\"asser}}, \bibinfo
  {author} {\bibfnamefont {C.}~\bibnamefont {Ederer}}, \bibinfo {author}
  {\bibfnamefont {M.}~\bibnamefont {F\"ahnle}}, \bibinfo {author}
  {\bibfnamefont {J.~M.}\ \bibnamefont {Sanchez}}, \ and\ \bibinfo {author}
  {\bibfnamefont {B.}~\bibnamefont {Meyer}},\ }\href@noop {} {\bibfield
  {journal} {\bibinfo  {journal} {Phys. Rev. B}\ }\textbf {\bibinfo {volume}
  {65}},\ \bibinfo {pages} {132104} (\bibinfo {year} {2002})}\BibitemShut
  {NoStop}%
\bibitem [{\citenamefont {Vanderbilt}(1985)}]{Vanderbilt85}%
  \BibitemOpen
  \bibfield  {author} {\bibinfo {author} {\bibfnamefont {D.}~\bibnamefont
  {Vanderbilt}},\ }\href@noop {} {\bibfield  {journal} {\bibinfo  {journal}
  {Phys. Rev. B}\ }\textbf {\bibinfo {volume} {32}},\ \bibinfo {pages} {8412}
  (\bibinfo {year} {1985})}\BibitemShut {NoStop}%
\bibitem [{\citenamefont {Jang}\ \emph {et~al.}(2009)\citenamefont {Jang},
  \citenamefont {Kim},\ and\ \citenamefont {Bhadeshia}}]{Jang09}%
  \BibitemOpen
  \bibfield  {author} {\bibinfo {author} {\bibfnamefont {J.~H.}\ \bibnamefont
  {Jang}}, \bibinfo {author} {\bibfnamefont {I.~G.}\ \bibnamefont {Kim}}, \
  and\ \bibinfo {author} {\bibfnamefont {H.~K. D.~H.}\ \bibnamefont
  {Bhadeshia}},\ }\href@noop {} {\bibfield  {journal} {\bibinfo  {journal}
  {Computational Materials Science}\ }\textbf {\bibinfo {volume} {44}},\
  \bibinfo {pages} {1319 } (\bibinfo {year} {2009})}\BibitemShut {NoStop}%
\bibitem [{\citenamefont {Jang}\ \emph {et~al.}(2010)\citenamefont {Jang},
  \citenamefont {Kim},\ and\ \citenamefont {Bhadeshia}}]{Jang10}%
  \BibitemOpen
  \bibfield  {author} {\bibinfo {author} {\bibfnamefont {J.~H.}\ \bibnamefont
  {Jang}}, \bibinfo {author} {\bibfnamefont {I.~G.}\ \bibnamefont {Kim}}, \
  and\ \bibinfo {author} {\bibfnamefont {H.~K. D.~H.}\ \bibnamefont
  {Bhadeshia}},\ }\href@noop {} {\bibfield  {journal} {\bibinfo  {journal}
  {Scripta Materialia}\ }\textbf {\bibinfo {volume} {63}},\ \bibinfo {pages}
  {121 } (\bibinfo {year} {2010})}\BibitemShut {NoStop}%
\bibitem [{\citenamefont {Henkelman}\ and\ \citenamefont
  {Jonsson}(2000)}]{Henkelman00}%
  \BibitemOpen
  \bibfield  {author} {\bibinfo {author} {\bibfnamefont {G.}~\bibnamefont
  {Henkelman}}\ and\ \bibinfo {author} {\bibfnamefont {H.}~\bibnamefont
  {Jonsson}},\ }\href@noop {} {\bibfield  {journal} {\bibinfo  {journal}
  {J.~Chem.~Phys.}\ }\textbf {\bibinfo {volume} {113}},\ \bibinfo {pages}
  {9978} (\bibinfo {year} {2000})}\BibitemShut {NoStop}%
\bibitem [{\citenamefont {Jiang}\ and\ \citenamefont {Carter}(2004)}]{Jiang04}%
  \BibitemOpen
  \bibfield  {author} {\bibinfo {author} {\bibfnamefont {D.~E.}\ \bibnamefont
  {Jiang}}\ and\ \bibinfo {author} {\bibfnamefont {E.~A.}\ \bibnamefont
  {Carter}},\ }\href@noop {} {\bibfield  {journal} {\bibinfo  {journal} {Phys.
  Rev. B}\ }\textbf {\bibinfo {volume} {70}},\ \bibinfo {pages} {064102}
  (\bibinfo {year} {2004})}\BibitemShut {NoStop}%
\bibitem [{VAS()}]{VASP_Note}%
  \BibitemOpen
  \href@noop {} {}\bibinfo {note} {For these LSDA-GGA calculations, in order to
  get the best comparison with refs [\onlinecite{Domain01}],
  [\onlinecite{Becquart07}] and [\onlinecite{Lau07}] we used the VASP
  code\cite{Kresse96,Kresse99} (version~4.6) with ultrasoft pseudopotentials
  and a plane wave basis (energy cut off 500eV) instead of the MBPP code. All
  other computational settings are as in section~\ref{sec_concentrated}. We
  made sure that the MBPP and VASP results for the four C$_2$ configurations
  agreed very closely for cubic $3\times 3\times 3$ supercells and then treated
  $4\times 4\times 4$ supercells only with VASP}\BibitemShut {NoStop}%
\bibitem [{\citenamefont {Kresse}\ and\ \citenamefont
  {Furthm{\"u}ller}(1996)}]{Kresse96}%
  \BibitemOpen
  \bibfield  {author} {\bibinfo {author} {\bibfnamefont {G.}~\bibnamefont
  {Kresse}}\ and\ \bibinfo {author} {\bibfnamefont {J.}~\bibnamefont
  {Furthm{\"u}ller}},\ }\href@noop {} {\bibfield  {journal} {\bibinfo
  {journal} {Computational Materials Science}\ }\textbf {\bibinfo {volume}
  {6}},\ \bibinfo {pages} {15} (\bibinfo {year} {1996})}\BibitemShut {NoStop}%
\bibitem [{\citenamefont {Kresse}\ and\ \citenamefont
  {Joubert}(1999)}]{Kresse99}%
  \BibitemOpen
  \bibfield  {author} {\bibinfo {author} {\bibfnamefont {G.}~\bibnamefont
  {Kresse}}\ and\ \bibinfo {author} {\bibfnamefont {D.}~\bibnamefont
  {Joubert}},\ }\href@noop {} {\bibfield  {journal} {\bibinfo  {journal}
  {\PRB}\ }\textbf {\bibinfo {volume} {59}},\ \bibinfo {pages} {1758} (\bibinfo
  {year} {1999})}\BibitemShut {NoStop}%
\bibitem [{\citenamefont {Lozovoi}\ and\ \citenamefont
  {Paxton}(2008)}]{Lozovoi08}%
  \BibitemOpen
  \bibfield  {author} {\bibinfo {author} {\bibfnamefont {A.~Y.}\ \bibnamefont
  {Lozovoi}}\ and\ \bibinfo {author} {\bibfnamefont {A.~T.}\ \bibnamefont
  {Paxton}},\ }\href@noop {} {\bibfield  {journal} {\bibinfo  {journal} {\PRB}\
  }\textbf {\bibinfo {volume} {77}},\ \bibinfo {pages} {165413} (\bibinfo
  {year} {2008})}\BibitemShut {NoStop}%
\bibitem [{\citenamefont {Tang}\ \emph {et~al.}(1996)\citenamefont {Tang},
  \citenamefont {Wang}, \citenamefont {Chan},\ and\ \citenamefont
  {Ho}}]{Tang96}%
  \BibitemOpen
  \bibfield  {author} {\bibinfo {author} {\bibfnamefont {M.~S.}\ \bibnamefont
  {Tang}}, \bibinfo {author} {\bibfnamefont {C.~Z.}\ \bibnamefont {Wang}},
  \bibinfo {author} {\bibfnamefont {C.~T.}\ \bibnamefont {Chan}}, \ and\
  \bibinfo {author} {\bibfnamefont {K.~M.}\ \bibnamefont {Ho}},\ }\href@noop {}
  {\bibfield  {journal} {\bibinfo  {journal} {Phys. Rev. B}\ }\textbf {\bibinfo
  {volume} {53}},\ \bibinfo {pages} {979} (\bibinfo {year} {1996})}\BibitemShut
  {NoStop}%
\bibitem [{\citenamefont {Haas}\ \emph {et~al.}(1998)\citenamefont {Haas},
  \citenamefont {Wang}, \citenamefont {F\"ahnle}, \citenamefont {Els\"asser},\
  and\ \citenamefont {Ho}}]{Haas98}%
  \BibitemOpen
  \bibfield  {author} {\bibinfo {author} {\bibfnamefont {H.}~\bibnamefont
  {Haas}}, \bibinfo {author} {\bibfnamefont {C.~Z.}\ \bibnamefont {Wang}},
  \bibinfo {author} {\bibfnamefont {M.}~\bibnamefont {F\"ahnle}}, \bibinfo
  {author} {\bibfnamefont {C.}~\bibnamefont {Els\"asser}}, \ and\ \bibinfo
  {author} {\bibfnamefont {K.~M.}\ \bibnamefont {Ho}},\ }\href@noop {}
  {\bibfield  {journal} {\bibinfo  {journal} {Phys. Rev. B}\ }\textbf {\bibinfo
  {volume} {57}},\ \bibinfo {pages} {1461} (\bibinfo {year}
  {1998})}\BibitemShut {NoStop}%
\bibitem [{\citenamefont {Nguyen-Manh}\ \emph {et~al.}(2000)\citenamefont
  {Nguyen-Manh}, \citenamefont {Pettifor},\ and\ \citenamefont
  {Vitek}}]{Pettifor00}%
  \BibitemOpen
  \bibfield  {author} {\bibinfo {author} {\bibfnamefont {D.}~\bibnamefont
  {Nguyen-Manh}}, \bibinfo {author} {\bibfnamefont {D.~G.}\ \bibnamefont
  {Pettifor}}, \ and\ \bibinfo {author} {\bibfnamefont {V.}~\bibnamefont
  {Vitek}},\ }\href@noop {} {\bibfield  {journal} {\bibinfo  {journal} {Phys.
  Rev. Lett.}\ }\textbf {\bibinfo {volume} {85}},\ \bibinfo {pages} {4136}
  (\bibinfo {year} {2000})}\BibitemShut {NoStop}%
\end{thebibliography}

\end{document}